\documentclass[a4paper,10pt,twocolumn]{article}  
\usepackage{graphicx,amsmath,amssymb,dsfont,bm}
\usepackage[utf8]{inputenc}
\usepackage{color}
\usepackage{enumitem}
\usepackage{boldline}
\usepackage{geometry}
 \usepackage{mathpazo}
\usepackage{authblk}
\usepackage[colorlinks=true,urlcolor=black,linkcolor=blue,citecolor=blue,bookmarks=false]{hyperref}
\usepackage{listings}
\usepackage{subfig}
\usepackage[normalem]{ulem}
\usepackage[switch]{lineno} 

\usepackage[superscript,biblabel,nomove]{cite}
\makeatletter
\renewcommand\@biblabel[1]{\textnormal{#1.~}}
\makeatletter

\usepackage{multirow}
\usepackage[percent]{overpic}
\usepackage{balance} 
\usepackage{helvet}
\usepackage{cuted}
\usepackage{adjustbox}
\usepackage{overpic} 
\usepackage{sectsty}
\usepackage{titlesec}
\titleformat*{\section}{\fontsize{12}{15}\sffamily\bfseries}
\titleformat*{\subsection}{\fontsize{10}{0}\sffamily\bfseries}
\titlespacing*{\section}{0pt}{12pt plus 4pt minus 2pt}{0pt plus 2pt minus 2pt}
\titlespacing*{\subsection}{0pt}{12pt plus 4pt minus 2pt}{0pt plus 2pt minus 2pt}

\usepackage{color} 
\usepackage{soul} 
\setstcolor{red} 
\makeatletter 
	\def\SOUL@ulthickness{2pt} 
\makeatletter

\title{\sffamily\bfseries\raggedright\huge How much longer do you have to drive than the crow has to fly?}
\author[*]{\sffamily\bfseries\raggedright  Shanshan Wang}
\author[$$]{\sffamily\bfseries\raggedright  Henrik M. Bette}
\author[$$]{\sffamily\bfseries\raggedright  Michael Schreckenberg}
\author[$$]{\sffamily\bfseries\raggedright  Thomas Guhr}

\affil[$$]{\sffamily Faculty of Physics, University of Duisburg-Essen, Duisburg, Germany}
\affil[*]{\sffamily Corresponding author: shanshan.wang@uni-due.de}
\affil[$$]{\sf \today}
\date{}

\begin{document}
\maketitle

\renewcommand{\abstractname}{}
\begin{strip}
\vspace*{-2cm}
\begin{abstract} 
\sffamily

When traveling by car from one location to another, our route is
constrained by the road network. The network distance between the two
locations is generally longer than the geodetic distance as the crow
flies. We report a systematic relation between the statistical
properties of these two distances. Empirically, we find a robust scaling
between network and geodetic distance distributions for a variety
of large motorway networks. A simple consequence is that we typically
have to drive $1.3\pm0.1$ times longer than the crow flies. This
scaling is not present in standard random networks; rather, it
requires non-random adjacency. We develop a set of rules to build a
realistic motorway network, also consistent with the above scaling. We
hypothesize that the scaling reflects a compromise between two
societal needs: high efficiency and accessibility on the one hand, and
limitation of costs and other burdens on the other.

\vspace*{0.55cm}
\end{abstract}
\end{strip}


\noindent
The development of a transportation network is an iterative process driven by the need for easy accessibility. In urban areas, streets are built to connect locations already in place, but the site selection for hospitals, shops, warehouses, etc., is influenced by the existing street network~\cite{Hansen1959,Geurs2004,Saif2019}. This also applies, \textit{mutatis mutandis}, to the construction of the motorway networks which began about one hundred years ago or later, depending on the country, after a certain level of industrialization had been reached. But the socio-economic benefits~\cite{Meersman2017,Rikalovic2014,Church2009} of a transportation network in a given country cannot grow indefinitely with the network's further enlargement. Construction of new motorways is costly and criticized in modern societies for environmental and other reasons. There is a kind of diminishing marginal utility.

The question arises as to how to quantitatively characterize this trade-off between conflicting interests. Of course, a general answer must involve a large variety of aspects, ranging from engineering and geographical matters to economic and demographic considerations to environmental and political issues. Here, we want to contribute to an answer by putting forward an approach that is based only on the motorway network itself, more specifically, on statistical properties of two kinds of distances.

To measure accessibility in a transportation network, a natural observable is the distance between locations~\cite{Hansen1959,Thornton2011}. The distance in a transportation network is measured by searching the shortest path on the network between two locations. This defines the network distance, which is distinct from the Euclidean and the geodetic distances between the same locations. For urban areas or, more generally, for a moderate extension of the network, the Euclidean distance suffices, but it should be replaced by the geodetic distance if the curvature of the Earth becomes relevant~\cite{Barthelemy2011}. These two distances ignore changes in elevation. The network distance includes curvature as well as elevation effects; the latter should be relatively small for the motorway networks in most countries.

Comparing the network and the Euclidean or geodetic distances gives information on accessibility in a transportation network. Various studies have been put forward. Correlation and regression analyses~\cite{Boscoe2012,Chen2021,Levinson2009,Levinson2012} were employed to quantify the network efficiency. Improved search methods in spatial databases~\cite{Papadias2003} were proposed. The ratio of network and Euclidean distances is often referred to as detour index or circuity. Slope and straightness centrality are related quantities whose distributions were investigated in Refs.~\citen{Crucitti2006a,Crucitti2006b}. Importantly, these studies are devoted to transportation on smaller scales or in metropolitan areas. In the context of urban data, various kinds of scaling behaviour were identified, e.g., for urban spatial structures~\cite{Carvalho2004,Li2017}, urban supply networks~\cite{Kuhnert2006} and urban road networks~\cite{Masucci2015,Kalapala2006}. Scaling properties shared by the distributions of urban and cropland networks~\cite{Strano2017} were identified. The scaling behaviour is often related to network distances~\cite{Lammer2006,Strano2013} or to the detour index~\cite{Yang2018}.

The scaling that the majority of the previous studies focus on is between network and Euclidean distances mostly in urban regions, leaving aside the non-urban regions. Thus far, empirical information and comparisons of the network and geodetic distance distributions for motorway networks covering urban and non-urban regions have not been offered. Motorway traffic features high driving speed, high traffic flow, an absence of traffic signal control, etc., distinguishing it from traffic on urban or other transportation networks. Here, we have three goals: first, we present a thorough empirical analysis of network and geodetic distance distributions for a variety of countries and larger areas with different geographic, topological, economic and political conditions. Second, we identify a surprisingly uniform scaling between the two distributions with remarkable stability observed in different countries. Third, by comparing simulated networks, we present strong evidence that this scaling is due to the presence of guiding principles rather than randomness in the development of a motorway network. We substantiate this by providing a set of rules that mimics realistic motorway networks, which is related to but different from other models as in Refs.~\citen{Yang1998, Jiang2004,Barthelemy2008,Barthelemy2011}.

\section*{Results}
\subsection*{Data analysis and empirical results}
\label{sec2}

\begin{figure*}[h!]
\begin{center}
\includegraphics[width=0.98\textwidth]{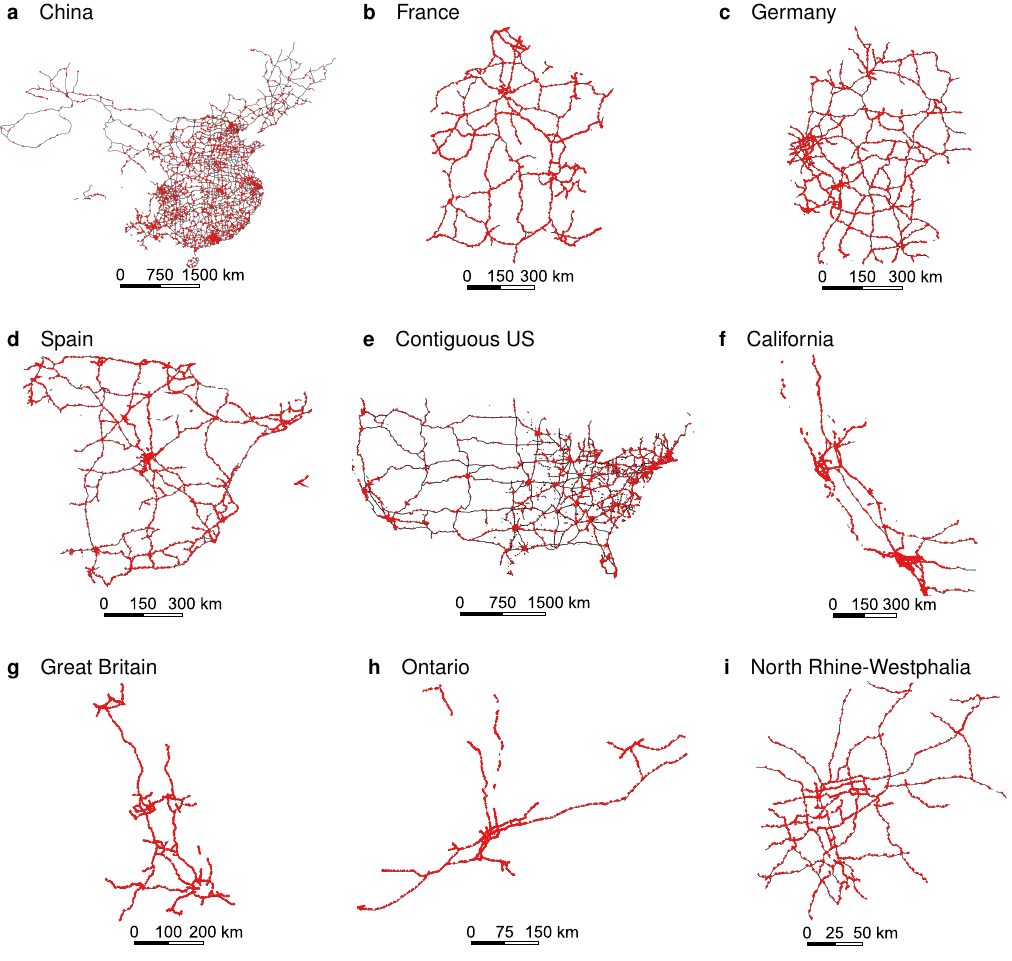}
\caption{Motorway networks (black lines) in China ({\textbf a}), France ({\textbf b}), Germany ({\textbf c}), Spain ({\textbf d}), the contiguous United States of America ({\textbf e}), California ({\textbf f}), Great Britain ({\textbf g}), Ontario ({\textbf h}) and North Rhine-Westphalia ({\textbf i}), respectively, with the locations (red dots) used to calculate distances. Motorway network data provided by OpenStreetMap (OSM) \copyright~OpenStreetMap contributors~\cite{osm, osmcopyright}. Maps developed with QGIS 3.4~\cite{qgis}.}
\label{fig1}
\end{center}
\end{figure*}

\begin{figure*}[h!]
\begin{center}
\includegraphics[width=0.98\textwidth]{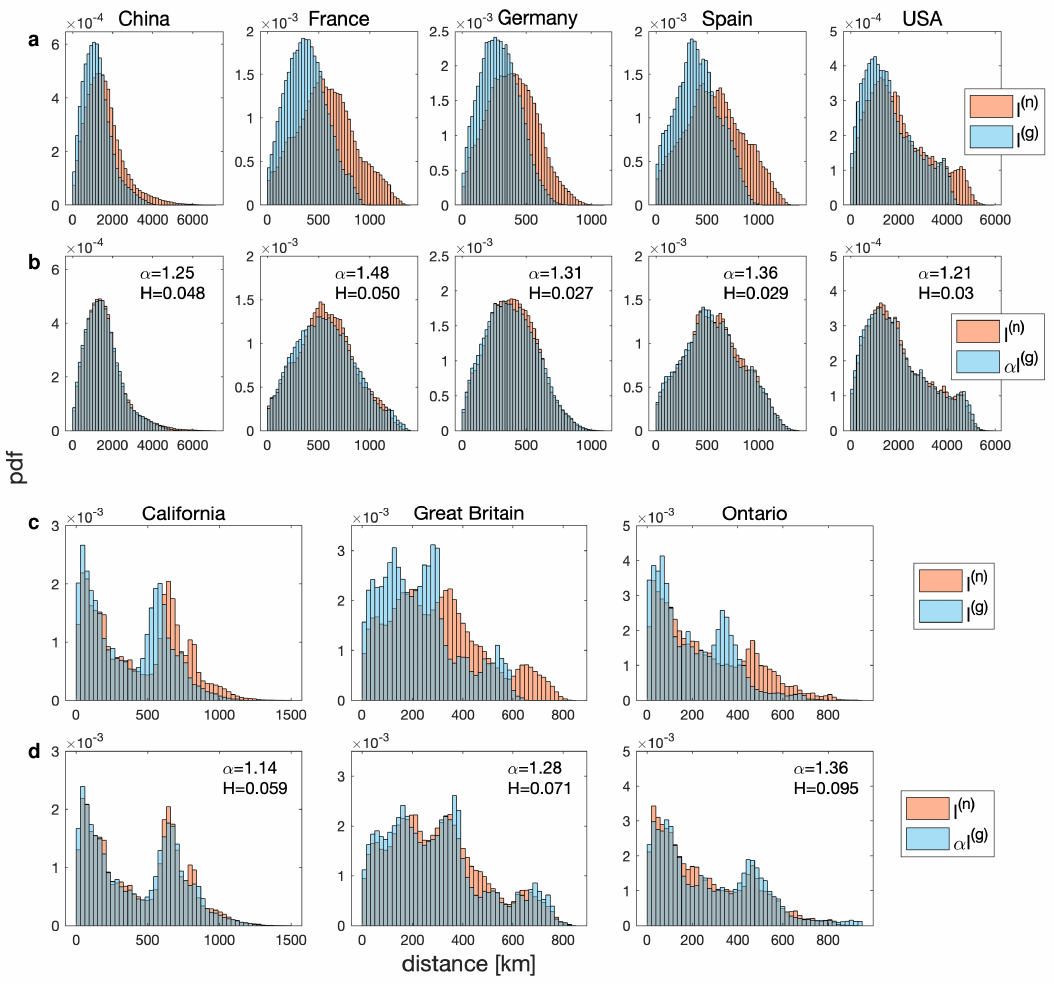}
\caption{Distributions or pdfs $p^\text{(n)}(l)$ and $p^\text{(g)}(l)$ of network and geodetic distances for the eight motorway networks in Fig.~\ref{fig1}, before ({\textbf a}, {\textbf c}) and after ({\textbf b}, {\textbf d}) scaling. Scaling factors $\alpha$ and Hellinger distances $H$ are given in the subfigures of {\textbf b} and {\textbf d}. Grey indicates overlap of two distributions.}
\label{fig2}
\end{center}
\end{figure*}

\begin{figure*}[tb]
\begin{center}
\includegraphics[width=0.98\textwidth]{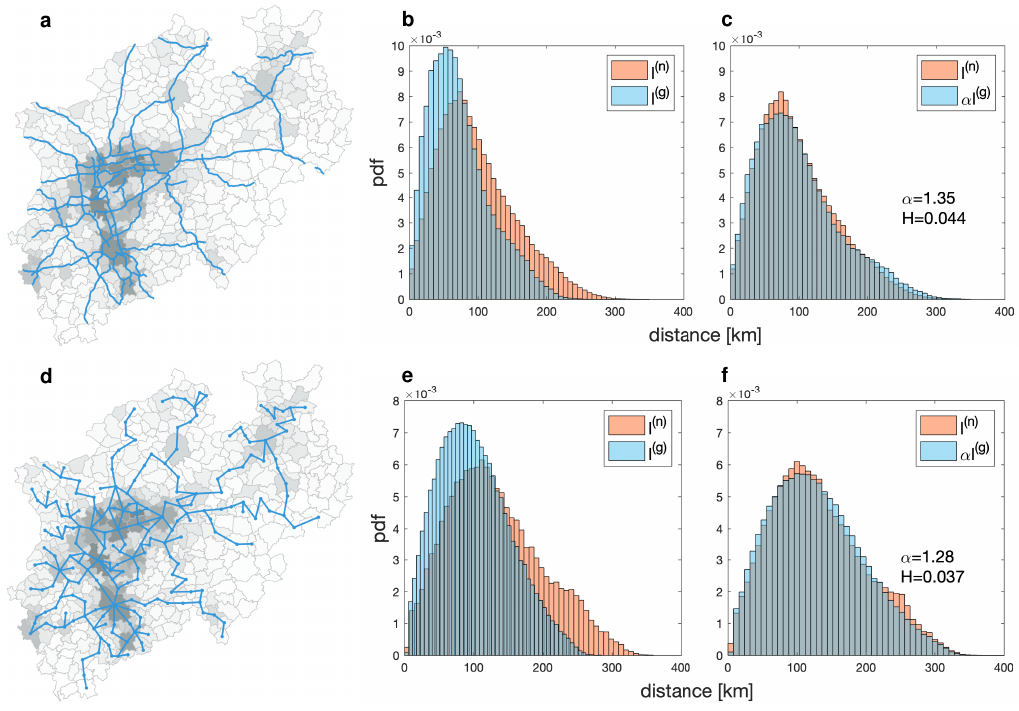}
\caption{{\textbf a}, real motorway network and {\textbf d}, constructed region motorway network for North Rhine-Westphalia. Distributions $p^\text{(n)}(l)$ and $p^\text{(g)}(l)$ of network and geodetic distances before ({\textbf b} and {\textbf e}) and after ({\textbf c} and {\textbf f}) scaling. {\textbf b} and {\textbf c} are for the real motorway network. {\textbf e} and {\textbf f} are for the constructed region motorway network. In {\textbf a} and {\textbf d}, the underlying grey colour indicates population density; the darker the colour, the higher the population density. Data on motorway networks (blue lines) and region boundaries (light grey lines) provided by OpenStreetMap (OSM) \copyright~OpenStreetMap contributors~\cite{osm,osmcopyright}. Population density data, licensed under BY-2.0~\cite{licence}, provided by \copyright~Statistische \"Amter des Bundes und der L\"ander, Germany~\cite{SABL}. Maps in {\textbf a} and {\textbf d} developed with Python~\cite{python}. 
}
\label{fig3}
\end{center}
\end{figure*}

\begin{figure*}[tb!]
\begin{center}
\includegraphics[width=0.98\textwidth]{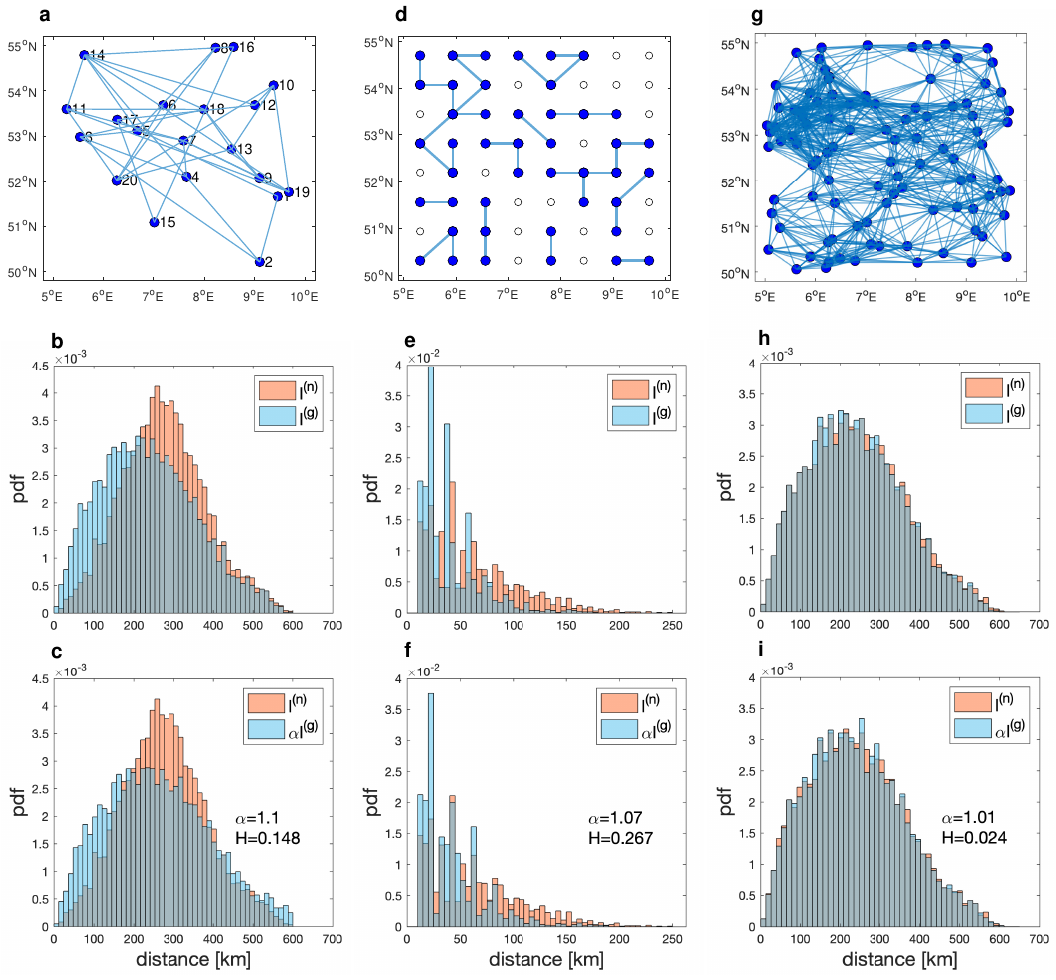}
\caption{{\textbf a}, fully random networks with $n=20$ locations and connection fractions $f=0.2$. {\textbf b} and {\textbf c}, distributions $p^\text{(n)}(l)$ and $p^\text{(g)}(l)$ of network and geodetic distances before ({\textbf b}) and after ({\textbf c}) scaling for fully random networks with $n=100$ locations and $f=0.2$. {\textbf d}, random grid networks with 64 locations on an $8\times 8$ grid and $f=0.2$. {\textbf e} and {\textbf f}, distributions $p^\text{(n)}(l)$ and $p^\text{(g)}(l)$ of network and geodetic distances before ({\textbf e}) and after ({\textbf f}) scaling for random grid networks with 900 locations on a $30\times 30$ grid and $f=0.2$. {\textbf g}, random geometric networks with $n=100$ locations and connection fractions $f=0.2$. {\textbf h} and {\textbf i}, distributions $p^\text{(n)}(l)$ and $p^\text{(g)}(l)$ of network and geodetic distances before ({\textbf h}) and after ({\textbf i}) scaling for random geometric networks with $n=100$ and $f=0.2$.}
\label{fig4}
\end{center}
\vspace*{-0.3cm}
\end{figure*}

\begin{figure*}[tb]
\begin{center}
\includegraphics[width=0.98\textwidth]{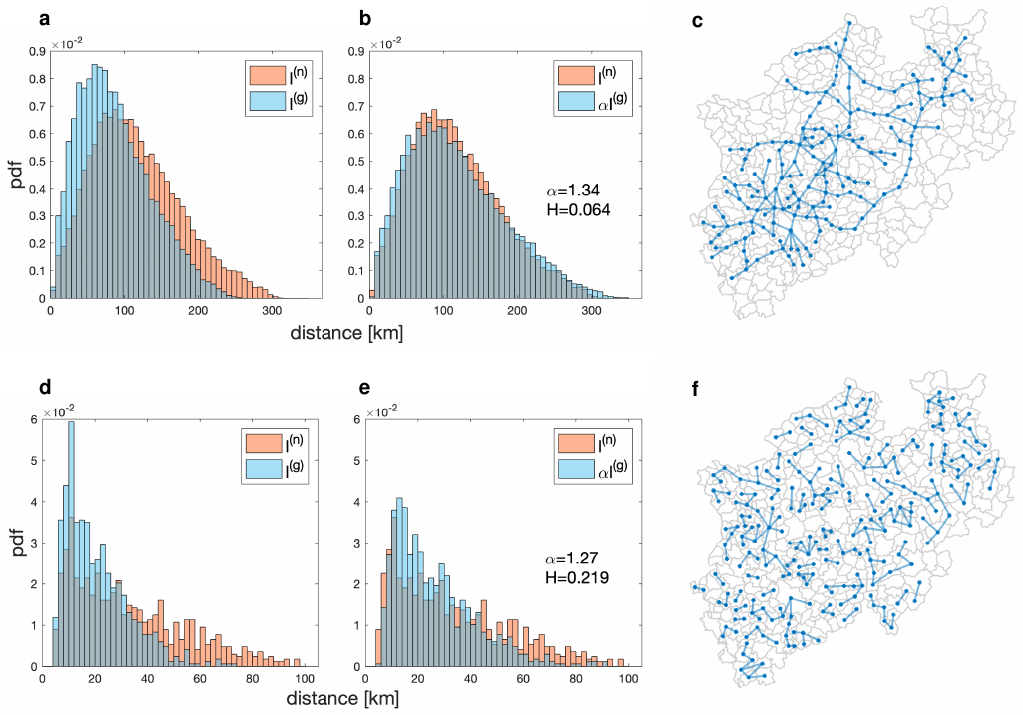}
\caption{Distributions or pdfs $p^\text{(n)}(l)$ and $p^\text{(g)}(l)$ of network and geodetic distances before ({\textbf a} and {\textbf d}) and after ({\textbf b} and {\textbf e}) scaling. {\textbf a} and {\textbf b} are for the partially random motorway network ({\textbf c}). {\textbf d} and {\textbf e} are for the fully random motorway network ({\textbf f}). The two networks ({\textbf c} and {\textbf f}) are distributed in the North Rhine-Westphalia region with connection probabilities $f=0.2$. Maps in {\textbf c} and {\textbf f} developed with Matlab~\cite{matlab}. }
\label{fig5}
\end{center}
\end{figure*}

To begin with, we choose the motorway networks in China, France, Germany, Spain and the contiguous part of the United States of America, i.e.~the US excluding Alaska and Hawaii, see Figs.~\ref{fig1}a-e. All motorway network data are provided by OpenStreetMap (OSM). For each network, we select around 2,000 locations on motorways. To account for network connectivity, we work out the network distance $l^\text{(n)}$ and the corresponding geodetic distance $l^\text{(g)}$ between each pair of locations if there is a path between them. The empirical results for the two probability density functions (pdf) or distributions $p^\text{(n)}(l)$ and $p^\text{(g)}(l)$ are shown in Fig.~\ref{fig2}a. When the distances appear as arguments, we drop the upper indices g and n. When we compare the different motorway networks, the scales and shapes differ as is to be expected. However, when we compare the two distributions for a given network, we find that the scaling property
\begin{equation}
p^\text{(g)}(l) = \alpha p^\text{(n)}(\alpha l) \ 
\label{eq1}
\end{equation}
is realized in a good approximation. The scaling factors $\alpha$ are empirically determined by minimizing the residual sum of the squares $(p^\text{(g)}(l)- \alpha p^\text{(n)}(\alpha l))^2$. As displayed in Fig.~\ref{fig2}b, the distributions almost fully agree after rescaling $p^\text{(n)}$. We obtain values of $\alpha$ between 1.2 and 1.5. For comparison, Ref.~\citen{Aldous2013} found a factor of 1.18 between network and Euclidean distances for US inter-city road networks by considering cities as nodes, which is related to but different from the data analyzed here. To quantify the similarity of the distribution pairs after rescaling, we use the Hellinger distance $H$~\cite{Hellinger1909}, see Methods. It satisfies the property $0 \leq H \leq 1$, and the better the agreement, the smaller it is. The empirical Hellinger distances in Fig.~\ref{fig2} are quite small, confirming the visual impression. The scaling~\eqref{eq1} implies $\langle l^k\rangle^\text{(n)}=\alpha^k\langle l^k\rangle^\text{(g)}$ for the $k$-th moments of the distributions, see Methods. For $k=1$, we obtain that the mean network distance is $\alpha$ times longer than the mean geodetic distance. Some features of the distributions can be approximatively understood with a simple analytical model, see Sec. S6 in SI.

Is the scaling property due to the relatively homogeneous structure of the motorway networks we examined? This is not so, as the analysis of the networks in Great Britain, i.e. the United Kingdom excluding Northern Ireland, California, USA, and Ontario, Canada, see Figs.~\ref{fig1}f-h, reveals. The topology of the networks is bimodal for California and Ontario, and even trimodal for Great Britain, resulting in distributions very different from the previous ones, as depicted in Fig.~\ref{fig2}c. Remarkably, the scaling property is still present, see Fig.~\ref{fig2}d. We obtain values of $\alpha$ between 1.1 and 1.4, slightly lower than the ones above because the network and geodetic distances between locations in different centres of the multimodal networks tend to be close. Despite the rich structures of the distributions, the Hellinger distances $H$ are small, see Fig.~\ref{fig2}. Hence, the scaling behaviour is, at least in the cases considered here, independent of the network topologies.

Does the scaling property require a relatively large motorway network? Not surprisingly, there is such a tendency, but in Sec. S2 of the Supplementary Information (SI), we present an analysis of the motorway networks in the 16 German states which reveals a remarkable robustness. Even the smaller states show approximate scaling, but typically with larger values of $\alpha$ and $H$.

Is the scaling behaviour robust when we modify the randomly chosen set of locations? To study this, we take a closer look at North Rhine-Westphalia (Nordrhein-Westfalen), see Fig.~\ref{fig1}i, the most populous German state, with its large motorway network. As Figs.~\ref{fig3}b, c show, the scaling is very well developed. We now randomly select other sets of 2,000 locations each and find only slight changes in the scaling factor $\alpha$ and Hellinger distance $H$. We also vary the number of locations in the chosen sets, up to 10,000, and do not see a significant change in $\alpha$ and $H$ either, see Sec. S3 in SI. Hence, the scaling is no coincidence; rather, it is a robust feature of large motorway networks.

\subsection*{Types of networks and scaling}
\label{sec3}

Which properties must a motorway network have to be realistic --- in particular, to exhibit scaling? We tackle this question by studying network models. A real motorway network connects locations. A network model consists of nodes and edges to which we in the present context refer as locations and motorways, respectively. We will work out the network and geodetic distances between the locations. We choose an area of approximately $550~\text{km}\times 550~\text{km}$, roughly corresponding to the sizes of the European countries and areas analyzed empirically. Let us begin with the simplest network, a fully random one, by randomly selecting $n$ locations, such that there are $N=n(n-1)/2$ possible motorways between them. Given $n$ locations and a connection fraction $f$ we randomly choose $fN$ motorways, see Fig.~\ref{fig4}a, for example. If two locations are connected by a path in the motorway network, we work out the geodetic and the shortest possible network distances. Figures~\ref{fig4}b, c display the distributions of the two kinds of distances before and after scaling, respectively, taking $f=0.2$ and $n=100$ as an example. As seen in Sec. S4.1 of SI, the corresponding distributions for $n=100$ strongly depend on the value of $f$. For smaller $f\leq 0.3$, the shapes are so different that scaling is absent, as revealed by the large Hellinger distances $H>0.1$. For $f\geq 0.4$ onwards, the distributions begin to coincide trivially with $\alpha=1$. This behaviour does not match the empirical findings. In the real world, a motorway connecting two locations would not avoid a location in between and close by. Thus, the fully random motorway network contains a high number of unrealistic motorways, which alters the statistical features. We infer that in a better motorway model, neighbouring locations ought to be connected. A good model with that feature is a random grid network, see Fig.~\ref{fig4}d, for instance. In the above specified area, we choose a $30 \times 30$ grid of locations at regular intervals, and only allow motorways connecting adjacent locations in any direction, including diagonally~\cite{Sohouenou2020}. Motorways in this model do not cross. According to a fraction $f$, we then select locations to be connected. When $f=0.2$, the scaling behaviour for distance distributions is absent for $H>0.1$ in Figs.~\ref{fig4}e, f. As shown in Sec. S4.2 in SI, the motorway network consists of unjoined parts which grow together beyond $f=0.3$ or so. Apart from strong discrepancies for very small $f$, the distributions for $f>0.3$ show some scaling with higher values of $\alpha$ and differing in the details, but corroborating the above reasoning. Furthermore, we examine random spatial networks with locally finite configurations, including random geometric networks and $K$-neighbour networks based on randomly selecting $n$ locations~\cite{Aldous2013,Aldous2010,Hirsch2015}, see Secs. S4.3 and S4.4, respectively. For the former, we connect two locations if their geodetic distance does not exceed a threshold. For the latter, we connect two locations if a location belongs to the $K$ closest neighbours of the other. The two kinds of networks are obviously local regarding the connections, see a random geometric network with $f=0.2$ in Fig.~\ref{fig4}g. The case of $K$-neighbour networks is analogous. The corresponding distributions of network and geodetic distances in Fig.~\ref{fig4}h are indistinguishable, resulting in a scaling factor of $\alpha=1.01$, see Fig.~\ref{fig4}i. This considerably differs from our above empirical findings for real motorway networks, which demonstrates that the two kinds of spatial networks are unable to describe the realistic motorway networks. The locality of network connections appears  not to be the decisive feature. 

Guided by these observations, we now set up a model that realistically mimics features of the North Rhine-Westphalia motorway network. In a real motorway network, see Fig.~\ref{fig1}i, the intersections are the nodes, but it would be insufficient to only consider those as locations. In the empirical analysis, any point on the motorway network can be the origin or destination of a journey. Moreover, the model can only be realistic if it is capable of realistically developing a motorway network. A real motorway network connects cities, districts, municipalities, etc., see Fig.~\ref{fig3}a, to which we refer as regions. They will serve as possible locations, but not all of them will be connected. The importance of a region largely depends on the number of inhabitants. The North Rhine-Westphalia motorway network has no unjoined pieces, prompting us to require that only adjacent regions be connected. To promote accessibility, more populous regions are connected by motorways, even if they are not adjacent. Hence, less populous regions in between become connected, but not all regions, adjacent or not, are connected by motorways. In North Rhine-Westphalia, most motorways are in the most populous area, the Rhine-Ruhr region. The challenge is to specify the regions using published data, and then to convert the existing motorway network to a model network with nodes placed in the centres of these 396 regions, see Fig.~\ref{fig3}d. We connect the regions by hand to obtain the best possible match with the real motorway network. When we compare the distributions of network and geodetic distances of the real and the region motorway networks, we find a sufficient match and some differences in the details, see Figs.~\ref{fig3}b, c, e and f. The scaling factors and Hellinger distances are $\alpha=1.35$, $H=0.044$ and $\alpha=1.28$, $H=0.037$, respectively. It is very important that the region motorway network allows us to determine a connection fraction; we find $f=0.2214$ for North Rhine-Westphalia.

\subsection*{A realistic network model for motorways}
\label{sec4}

We can now provide rules for the realistic planning and construction of motorway networks. To the best of our knowledge, there is no such model in the literature. We put forward a partially random network model based on the above findings. The remaining randomness lies in the selections of regions and connecting motorways, allowing flexibility. We will then apply it to North Rhine-Westphalia. For a given connection fraction $f$, we construct the model network $G$ of $m$ motorways using the following procedure:
\begin{enumerate}[leftmargin=*]
\setlength\itemsep{-0em}
\item Construct a fully connected network $G_\text{all}$ of $n$ regions and $m_\text{all}$ motorways without crossings.
\item Randomly select a pair of regions with a selection probability $\omega_{ij}$ to be specified below.
\item Search the shortest path with length denoted by $s_{ij}$ between this pair $i$ and $j$ of regions in $G_\text{all}$, where $s_{ij}$ is the sum of the geodetic distances $l_{kl}^\text{(g)}$ of all the adjacent regions $k$ and $l$ connected by the path being considered. In the first application of this step, these resulting motorways between the adjacent regions $k$ and $l$ are taken as the first motorways in $G$. In the later reiterations, only those motorways which are not yet in $G$ are added to $G$.
\item Reward if a motorway already exists in $G$ by effectively shortening the corresponding $l_{kl}^\text{(g)}$ in $G_\text{all}$ according to $l_{kl}^\text{(g)} \leftarrow l_{kl}^\text{(g)}\varepsilon_{ij}$. The parameter $\varepsilon_{ij}$ is between 0 and 1 and will be specified below.
\item Repeat steps 2 to 4 until the number of motorways $m$ in $G$ reaches $m\geq \text{int\,}(fm_\text{all})$, where $\text{int\,}$ returns the integer closest to its argument.
\item For $m>\text{int\,}(fm_\text{all})$, randomly remove a motorway that connects a region which connects only to one other region. Repeat removing edges one by one until $m=\text{int\,}(fm_\text{all})$.
\end{enumerate}
We now apply this model to North Rhine-Westphalia. The $n=396$ regions are connected by $m_\text{all}=1084$ motorways to form $G_\text{all}$, see Sec. S1 in SI. We use the connection fraction $f=0.2214$ obtained above to calculate the number of motorways $m$ in the region network $G$, $m=\text{int\,}(fm_\text{all})=240$. As the population densities $P_i$ of the regions $i$ play a crucial role in the topology of the motorway network, we choose the selection probability 
\begin{equation}
\omega_{ij}=\frac{P_iP_j}{\sum_{i> j}P_iP_j } 
\label{eq43}
\end{equation}
which follows an exponential distribution. Paths are searched between regions $i$ and $j$ only if they are chosen according to $\omega_{ij}$, and not considered otherwise. It is also useful to relate the updating parameter $\varepsilon_{ij}$ to the population densities, 
\begin{equation}
\varepsilon_{ij}=\frac{\eta_{ij}}{\text{max}(\eta_{ij})} \quad \text{where} \quad
\eta_{ij}=\frac{\ln P_i \ln P_j}{\sum_{i> j}\ln P_i\ln P_j} \ 
\label{eq44}
\end{equation}
follows a log-normal distribution. For $\varepsilon_{ij}=0$, the distance $l_{kl}^{(g)}$ in the previously chosen path becomes zero, such that the shortest path for the next region pair must pass through the adjacent regions $k$ and $l$ instead of choosing a path shorter than the original $l_{kl}^{(g)}$. Thus setting $\varepsilon_{ij}=0$ results in a minimum spanning tree. On the contrary, letting $\varepsilon_{ij}=1$, a shorter path will replace the connection between $k$ and $l$ ignoring the significance of the region, which generates more loops in the network. Our setting in Eq.~\eqref{eq44} is a realistic compromise between the two extremes. The choice of $\varepsilon_{ij}$ favours the additional generation of paths in $G$ that run somewhat parallel to existing paths for densely populated regions while preventing this for sparsely populated regions. Such behaviour is observed in real motorway networks. 
This model captures salient features of the region motorway network for North Rhine-Westphalia, as we now demonstrate.

A crucial feature of our partially random motorway network is the adjacency, deeply rooted in the above rules, that produces fully connected networks rather than a network of unjoined pieces. For different connection fractions $f$, we present such modelled partially random motorway networks in Fig.~S12 of SI. For comparison, Fig.~S14 in SI depicts fully random motorway networks for the same connection fractions $f$, generated by randomly selecting motorways from the fully connected North Rhine-Westphalia region network, see Sec. S1 in SI, avoiding motorway crossings. We display the distributions $p^\text{(n)}(l)$ and $p^\text{(g)}(l)$ of network and geodetic distances in Figs.~S13 and S15 of SI for the partially and fully random motorway networks in Figs.~S12 and S14. In Fig.~\ref{fig5}, we compare the two kinds of networks for $f=0.2$. Their network structures as well as the corresponding distributions differ quite a bit. Unlike fully random networks and random grid networks, the partially random networks have distributions very similar to the empirical ones in the real North Rhine-Westphalia motorway network as borne out by lower Hellinger distances, especially when $f<0.4$. The corresponding scaling factors are close to our empirical results when $0.2\leq f \leq 0.3$. We infer that the very similar scaling properties corroborate that the above set of rules is capable of producing realistic models for motorway networks.

\section*{Discussion}
\label{sec5}

When developing a motorway network, two societal needs are in competition: accessibility and efficiency on the one hand, and cost savings and environmental protection on the other. In an attempt to determine criteria that help to characterize motorway networks in modern countries, we identified a new scaling property that relates the network and the geodetic distances in a remarkably stable manner. We confirm this in a variety of empirical analyses. The extracted scaling factors mean among other things that, on average, the network distance to be traveled is typically $1.3\pm0.1$ times longer than the geodetic distance. This scaling must reflect the aforementioned competition, but it can be analyzed on the basis of the motorway networks without additional input. Scaling is best realized in large motorway networks but, surprisingly, even smaller and less homogeneously distributed ones exhibit its onset quite clearly.

We showed that the scaling property is incompatible with simple structures as in fully random networks. Rather, the feature of adjacency is crucial, i.e.~real motorway networks develop in such a way that existing connections are most efficiently used. This observation led us to propose a new model: the partially random motorway network, in which motorways grow by connecting adjacent regions step by step. This ensures connectivity. We applied this model to the case of North Rhine-Westphalia, and showed that it reproduces the scaling found empirically very well for the correct connection fraction determined empirically.

In summary, we found a new universal scaling property in motorway networks empirically and, guided by its features, constructed a new, realistic model for such networks.

\section*{Methods}
\subsection*{Geodetic distance} 

The geodetic distance between two locations $i$ and $j$ is given by the haversine formula~\cite{Gade2010}
\begin{equation}
  l_{ij}^\text{(g)}=2R\text{arcsin}\left( \sqrt{\text{sin}^2\frac{\varphi_j-\varphi_i}{2}+\text{cos}\varphi_i\text{cos}\varphi_j\text{sin}^2\frac{\lambda_j-\lambda_i}{2}}\right) \ .
\label{eqM1}
\end{equation}
Here, $\varphi_i$ and $\lambda_i$ represent the latitude and longitude of location $i$, respectively. The Earth's radius $R$ is approximately $R=6371$ km.

\subsection*{Network distance} 
The network distance $l_{ij}^\text{(n)}$ of the shortest path between two locations $i$ and $j$ is measured by combining the tools Osmosis, OSMnx and NetworkX. Osmosis, a command line of Java applications, is used to filter the geospatial data for a motorway network. OSMnx is a Python package for reading the filtered geospatial data and identifying two given locations separately as an origin and a destination. NetworkX, also a Python package, is employed to search the shortest path between two locations with algorithms, e.g.~Dijkstra's algorithm~\cite{Dijkstra2022} in this study, and to calculate the length of this route only on the motorway network being examined. By exchanging the origin and the destination of two given locations, the distances in the magnitude of kilometers change very little. Hence we use the approximation $l_{ij}^\text{(n)}\approx l_{ji}^\text{(n)}$. According to the motorway network data provided by OpenStreetMap (OSM) \copyright~OpenStreetMap contributors~\cite{osm, osmcopyright}, a motorway connection consists of many, rather small pieces. The length of each edge is close to the geodetic distance $l_{ij}^\text{(g)}$ between the two adjacent nodes. Multiple paths may exist for a given pair of locations. The network distance of the shortest path between two locations minimizes the sum of geodetic distances along this path. We ignore the network distance if there is no connecting path between two locations. For the general case in our main text, we drop the subscript $ij$ from the distances $l^\text{(g)}$ and $l^\text{(n)}$.

\subsection*{Scaling property and moments}
The extra factor of $\alpha$ in front of $p^\text{(n)}(\alpha l)$ in the scaling law \eqref{eq1} follows from the very definition of a pdf. It is needed, for example, to ensure normalization. The moments with order $\kappa=0,1,2,\ldots$ of the distributions are
\begin{equation}
\langle l^\kappa\rangle^\text{(z)} = \int\limits_0^\infty p^\text{(z)}(l) l^\kappa dl \ , \quad \text{z}=\text{n},\text{g} \ .
\label{m2}
\end{equation}
The scaling property \eqref{eq1} for the distributions implies the scaling $\langle l^\kappa\rangle^\text{(n)} = \alpha^\kappa \langle l^\kappa\rangle^\text{(g)}$ for the moments with $\kappa=0,1,2,\ldots$. Thus, all centred moments and cumulants scale with $\alpha^\kappa$, too. In particular, the mean value $\langle l\rangle^\text{(z)}$ and the standard deviation scale with $\alpha$. Here, $\text{var}^\text{(z)} = \langle l^2\rangle^\text{(z)} - \langle l\rangle^{\text{(z)}\: 2}$ is the variance and $\text{std}^\text{(z)}=\sqrt{\text{var}^\text{(z)}}$ is the
standard deviation.

\subsection*{Hellinger distance} 
The Hellinger distance~\cite{Hellinger1909} of two distributions $q_1(x)$ and $q_2(x)$ on support $X$ is defined by 
\begin{eqnarray}
H&=&\sqrt{\frac{1}{2}\int\limits_X\left(\sqrt{q_1(x)}-\sqrt{q_2(x)}\right)^2dx}\nonumber\\
&=&\sqrt{1-\int\limits_X\sqrt{q_1(x)q_2(x)}dx} \ .
\label{eqM4}
\end{eqnarray}
By construction, it is zero for $q_1(x)=q_2(x)$. The Hellinger distance satisfies the property $0\leq H\leq 1$. If $H$ approaches zero, the two distributions exhibit a high similarity. In contrast, if $H$ tends towards one, they differ greatly.

\section*{Data availability}
All data supporting the findings in this work are available within the paper and its Supplementary Information. The raw data on motorways are provided by OpenStreetMap (\url{https://www.openstreetmap.org}) and can be directly downloaded from Geofabrik (\url{https://download.geofabrik.de}). The source data on population density is from Statistische \"Amter des Bundes und der L\"ander, Germany (\url{https://regionalatlas.statistikportal.de}). The empirical data points on motorways for calculation are available via GitHub at \url{https://github.com/shannwang/distancescaling}

\section*{Code availability}
The code used for empirical analyses and numerical simulations is available via GitHub at \url{https://github.com/shannwang/distancescaling}

\section*{Acknowledgements}

We gratefully acknowledge the support received from the German Research Foundation (DFG) as part of the project “Correlations and their Dynamics in Freeway Networks” (No. 418382724). We thank Andreas Schadschneider and Sebastian Gartzke for fruitful discussions.

\section*{Author contributions}

T.G. and M.S. proposed the research. S.W. conducted the empirical analysis. S.W. and T.G. did most of the writing. H.B. and S.W. developed the methods of modelling, with input from M.S. and T.G. All authors contributed equally to developing the methods of analysis and revising and reviewing the paper.

\section*{Competing interests}
The authors declare no competing interests. 


\end{document}


\maketitle

\noindent\rule{\textwidth}{1pt}
\vspace*{-1cm}
{\setlength{\parskip}{0pt plus 1pt} \tableofcontents}
\noindent\rule{\textwidth}{1pt}

\renewcommand{\thepage}{S\arabic{page}}
\renewcommand{\thesection}{S\arabic{section}}
\renewcommand{\thetable}{S\arabic{table}}
\renewcommand{\thefigure}{S\arabic{figure}}
\renewcommand{\theequation}{S\arabic{equation}}

\vspace*{2cm}
\section{Motorways in German states and region networks}

\begin{figure*}[htb!]
\begin{center}
\begin{overpic}[width=0.5\textwidth]{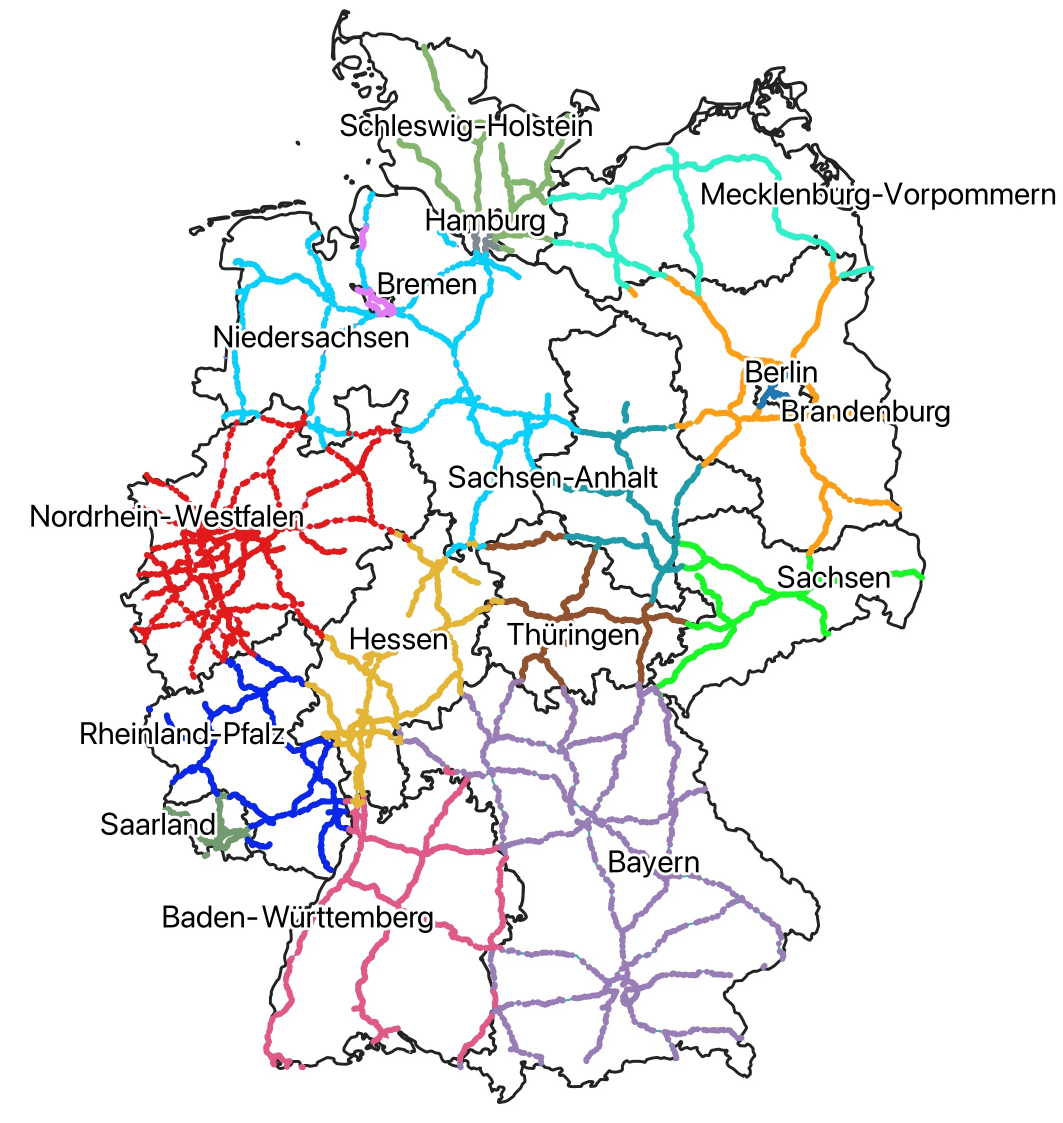}
\put(10,90){\bf \textsf a}
\end{overpic}
\begin{overpic}[width=0.49\textwidth]{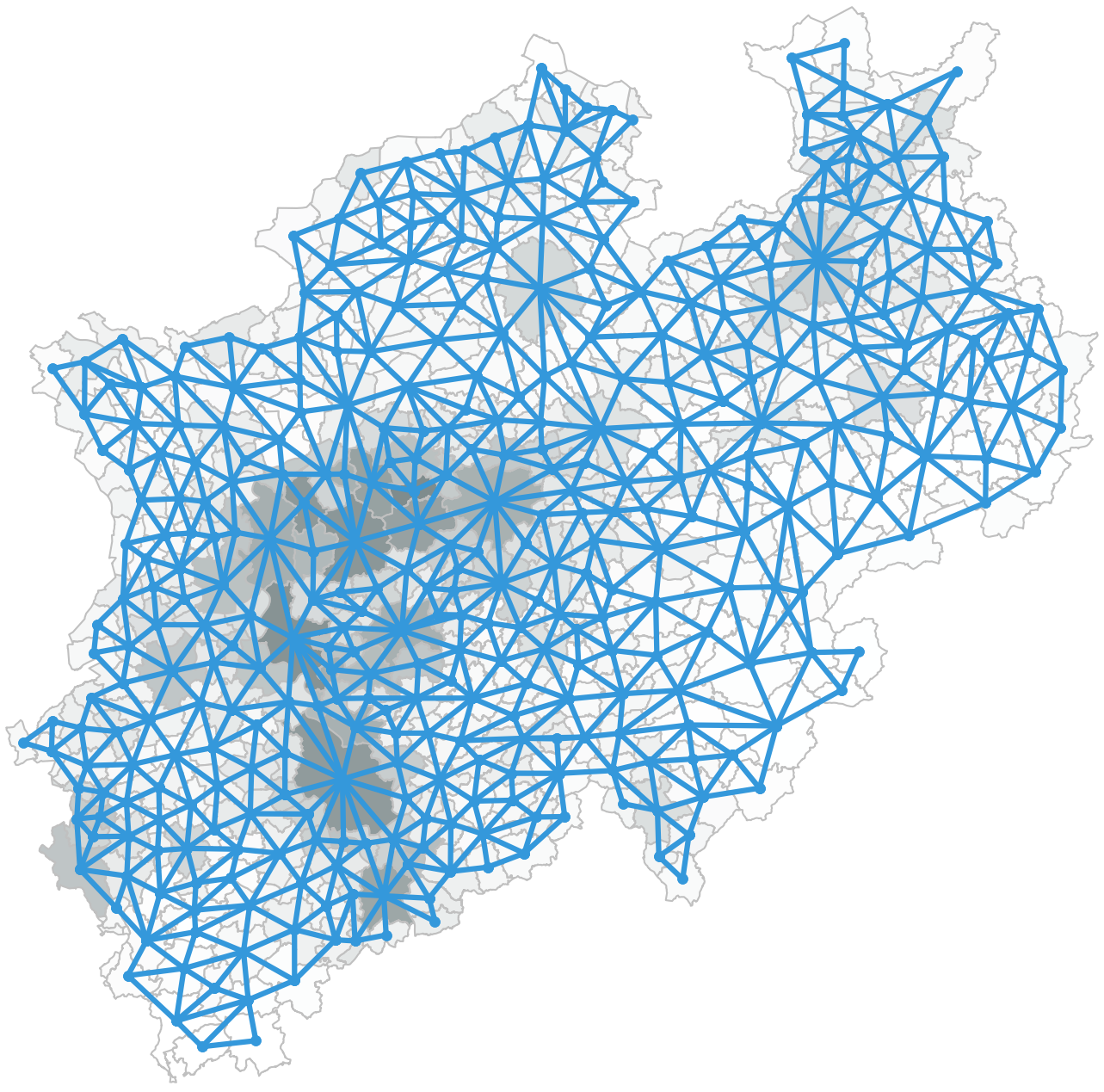}
\put(10,96){\bf \textsf b}
\end{overpic}
\caption{{\textbf a}, motorway networks in the 16 German states with locations used for distance computation. Data provided by OpenStreetMap (OSM) \copyright~OpenStreetMap contributors~\cite{osm,osmcopyright}. Map developed by QGIS 3.4~\cite{qgis}. {\textbf b}, fully connected region network for North Rhine-Westphalia. Grey with shadowing indicates population density; the darker the colour, the higher the population density. Light grey lines indicate region boundaries. Data provided by OpenStreetMap (OSM) \copyright~OpenStreetMap contributors~\cite{osm,osmcopyright}. Population density data licensed under BY-2.0~\cite{licence}, provided by \copyright~Statistische \"Amter des Bundes und der L\"ander, Germany~\cite{SABL}. Map developed with Python~\cite{python}.}
\label{figs1}
\end{center}
\end{figure*}

The German motorway network covers the 16 German states: North Rhine-Westphalia (Nordrhein-Westfalen), Bavaria (Bayern), Hesse (Hessen), Schleswig-Holstein, Baden-W\"urttemberg, Mecklenburg-Western Pomerania (Mecklenburg-Vorpommern), Lower Saxony (Niedersachsen), Brandenburg, Rhineland-Palatinate (Rheinland-Pfalz), Saxony-Anhalt (Sachsen-Anhalt), Saxony (Sachsen), Thuringia (Th\"uringen), Saarland, Hamburg, Berlin and Bremen. The locations of each of the 16 state motorway networks are marked with different colours, see Fig.~\ref{figs1}a. North Rhine-Westphalia is the most densely populated state in Germany, and thus has the densest motorway network. In our paper, we take the North Rhine-Westphalia motorway network as an example for empirical analysis as well as for network modelling. For the latter function, we separate this state into 396 regions, including cities, districts, municipalities, etc. Each region serves as a location, only some of which are connected by motorways. To model the network, we select motorways with a connection fraction. When the fraction is equal to one, we obtain a fully connected network $G_\mathrm{all}$ where each motorway connects two adjacent locations, see Fig.~\ref{figs1}b. With a given connection fraction, we pick up motorways among all motorways from the fully connected motorway network in Fig.~\ref{figs1}b.

\section{Network and geodetic distributions for the individual German states}

\begin{figure*}[htbp]
\begin{center}
\begin{overpic}[width=0.95\textwidth]{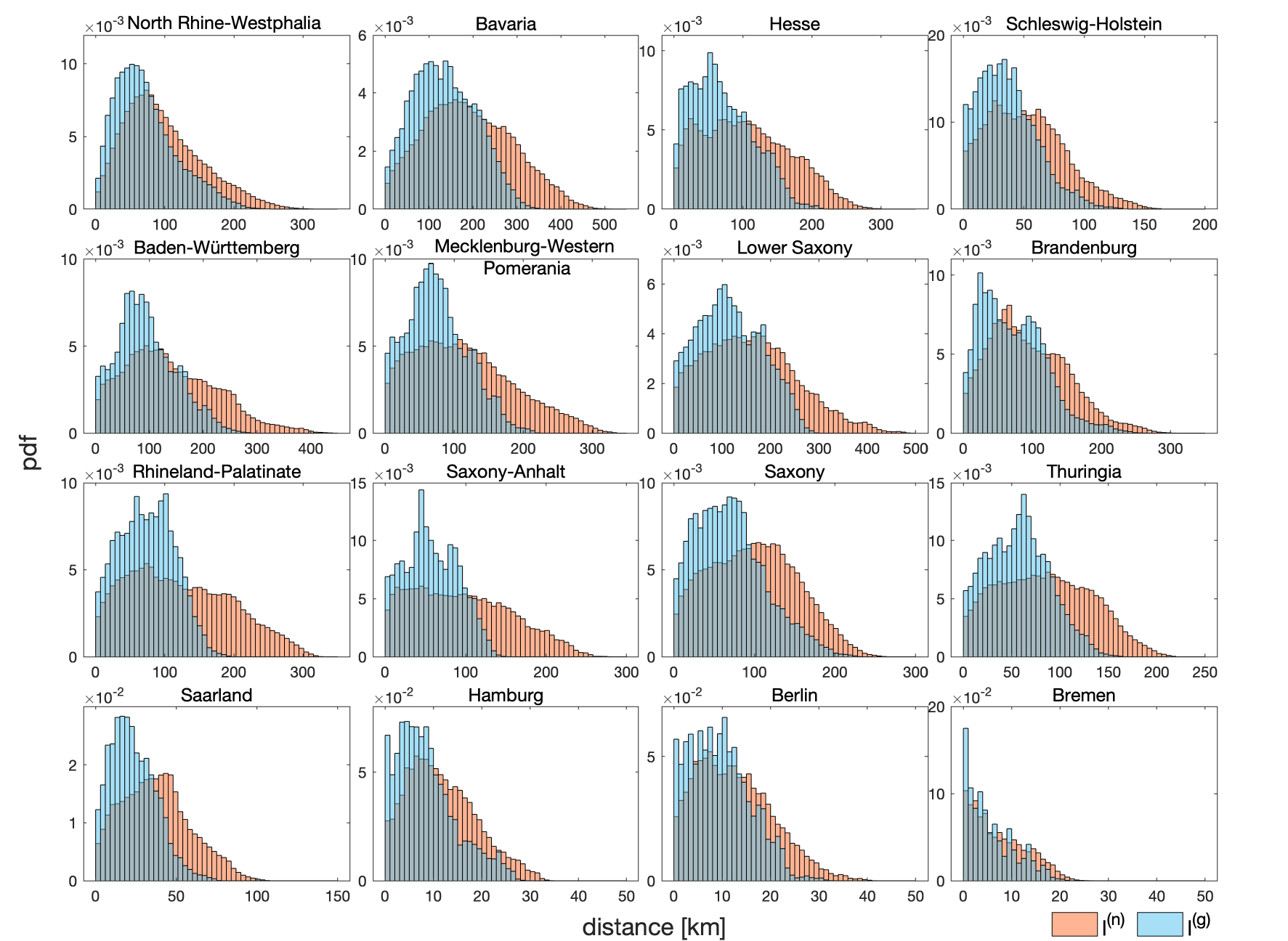}
\put(2,70){\bf \textsf a}
\end{overpic}
\\ \vspace{0.5cm}
\begin{overpic}[width=0.95\textwidth]{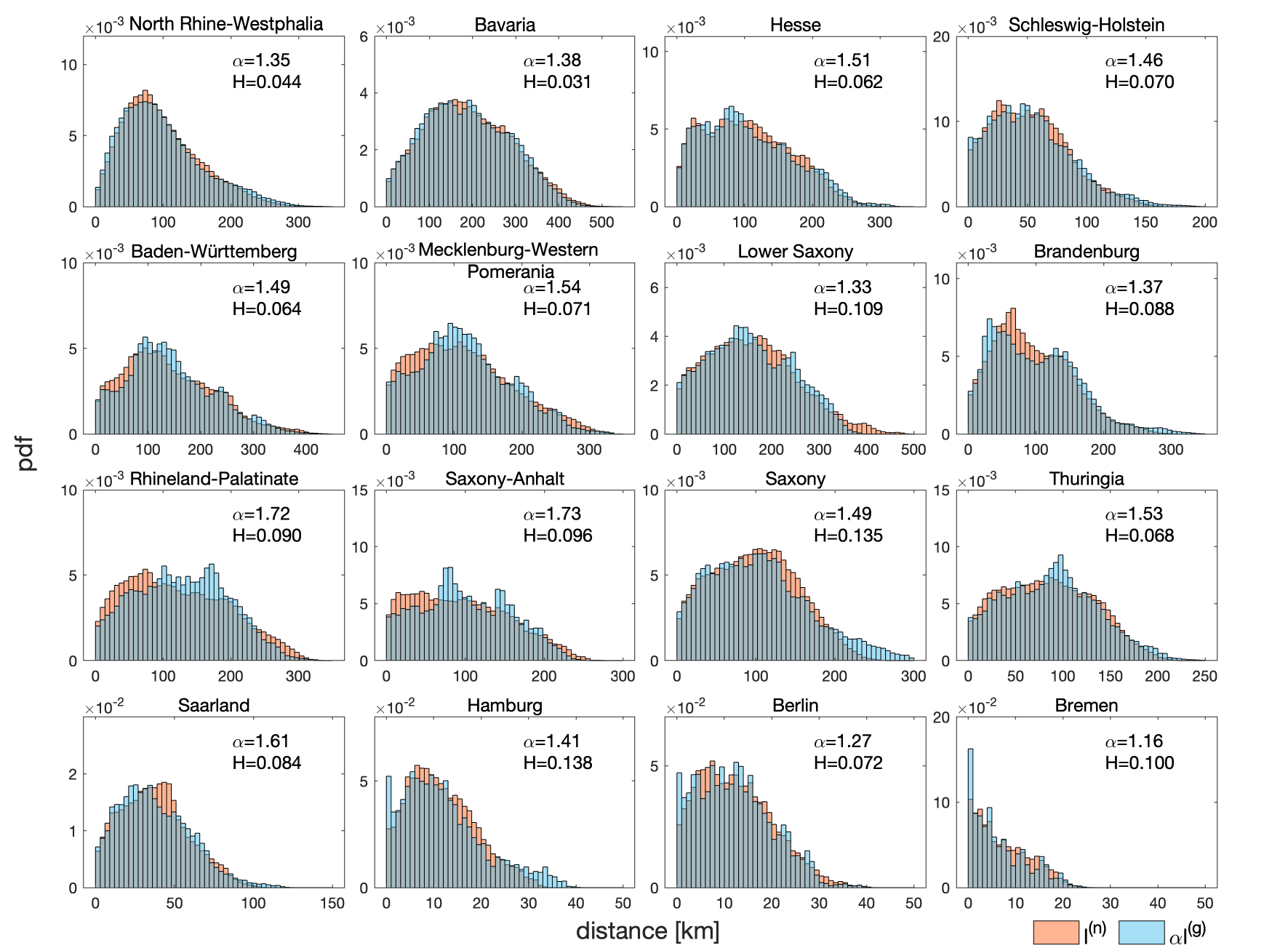}
\put(2,70){\bf \textsf b}
\end{overpic}
\caption{Distributions or pdfs $p^\text{(n)}(l)$ and $p^\text{(g)}(l)$ of network and geodetic distances for the motorway networks in the 16 German states, before ({\textbf a}) and after ({\textbf b}) scaling. The values of the scaling factor $\alpha$ and the Hellinger distance $H$ are given in the subfigures of {\textbf b}.}
\label{figs2}
\end{center}
\end{figure*}

Figure~\ref{figs2} displays the distributions of network and geodetic distances before and after scaling for the 16 state motorway networks in Germany. Let $H=0.1$ be a critical value for the Hellinger distance to distinguish a good from a poor similarity of two distributions. There are 13 motorway networks that exhibit the scaling behaviour very well. The distances in the state motorway networks are much smaller than for countries. But, remarkably, this does not affect the scaling behaviour. In particular, the scaling behaviour can be clearly observed even in the small Berlin and Bremen motorway networks. Hamburg, also a small network, does not exhibit the scaling behaviour so well, but approximate scaling can be seen with a higher scaling factor $\alpha$ and Hellinger distance $H$. Therefore, the scaling behaviour is rather robust, even for small networks.

\clearpage

\section{Robustness of the scaling behaviour}

\begin{figure*}[htb!]
\begin{center}
\hspace*{0.08cm}
\begin{overpic}[width=0.89\textwidth]{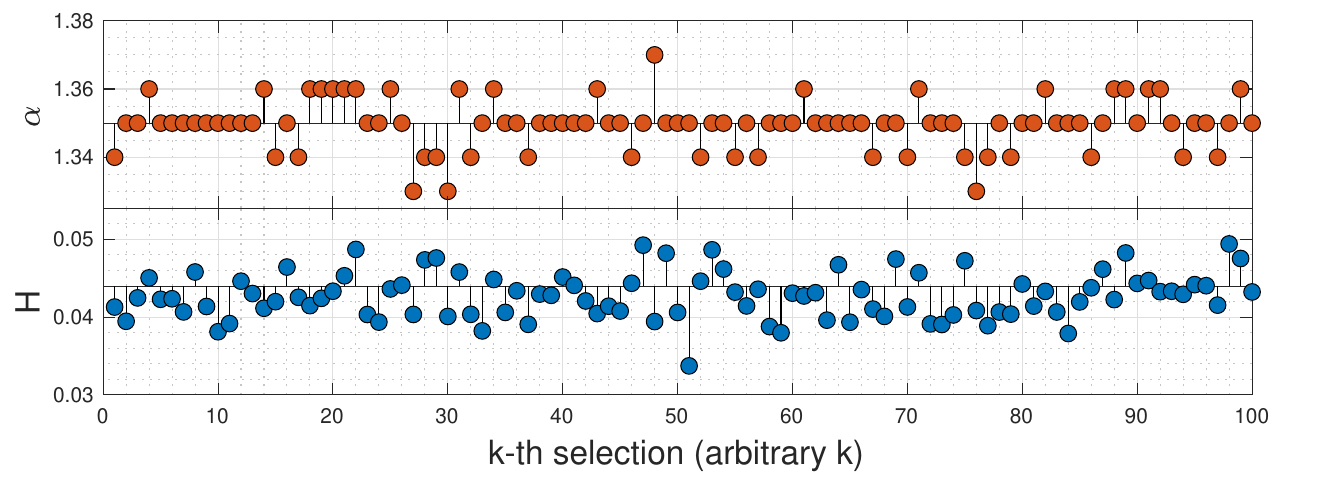}
\put(10,33){\bf \textsf a}
\put(10,18){\bf \textsf b}
\end{overpic}
\begin{overpic}[width=0.9\textwidth]{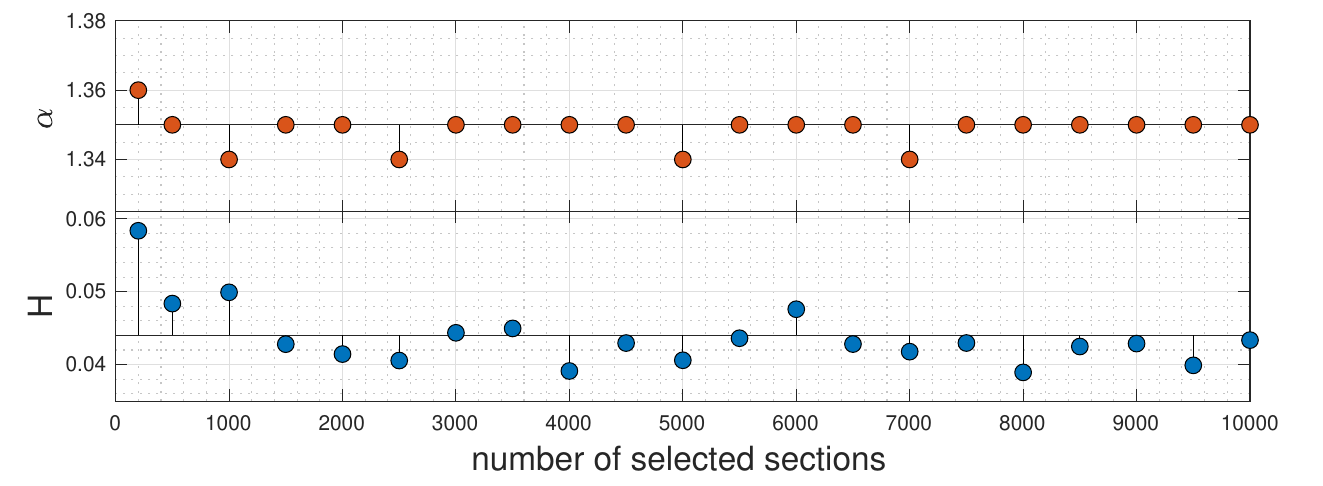}
\put(10,33.5){\bf \textsf c}
\put(12,18){\bf \textsf d}
\end{overpic}
\caption{Scaling factor $\alpha$ in {\textbf a} and corresponding Hellinger distances $H$ in {\textbf b} for 100 random selections of 2,000 motorway locations and for different numbers of randomly selected motorway locations in {\textbf c} and {\textbf d}.}
\label{figs3}
\end{center}
\end{figure*}

For North Rhine-Westphalia, uniformly selecting around 2,000 sections yields the scaling factor $\alpha=1.35$ and the Hellinger distance $H=0.044$, see Fig.~3b, c in the paper. We now randomly select 100 times exactly 2,000 sections from the North Rhine-Westphalia motorway network. The resulting scaling factors are shown in Fig.~\ref{figs3}a. Most are $\alpha=1.35$, with small fluctuations. Accordingly, the Hellinger distances in Fig.~\ref{figs3}b for most cases are below $H=0.044$ and only a few are above. The scaling is thus independent of the choice of locations. Furthermore, we enlarge the number of randomly selected locations from 200 to 10,000. The influence on the scaling factor is negligible as long as the number of sections selected is not too small, e.g.~larger than 500, see Figs.~\ref{figs3}c, d. Hence, the scaling is also independent of the number of locations. In other words, its stable value is reached quickly. We infer that the scaling is a robust empirical fact.

\vspace*{2cm}

\clearpage
\section{Random networks}

\subsection{Fully random networks}

\begin{figure*}[htbp]
\begin{center}
\includegraphics[width=\textwidth]{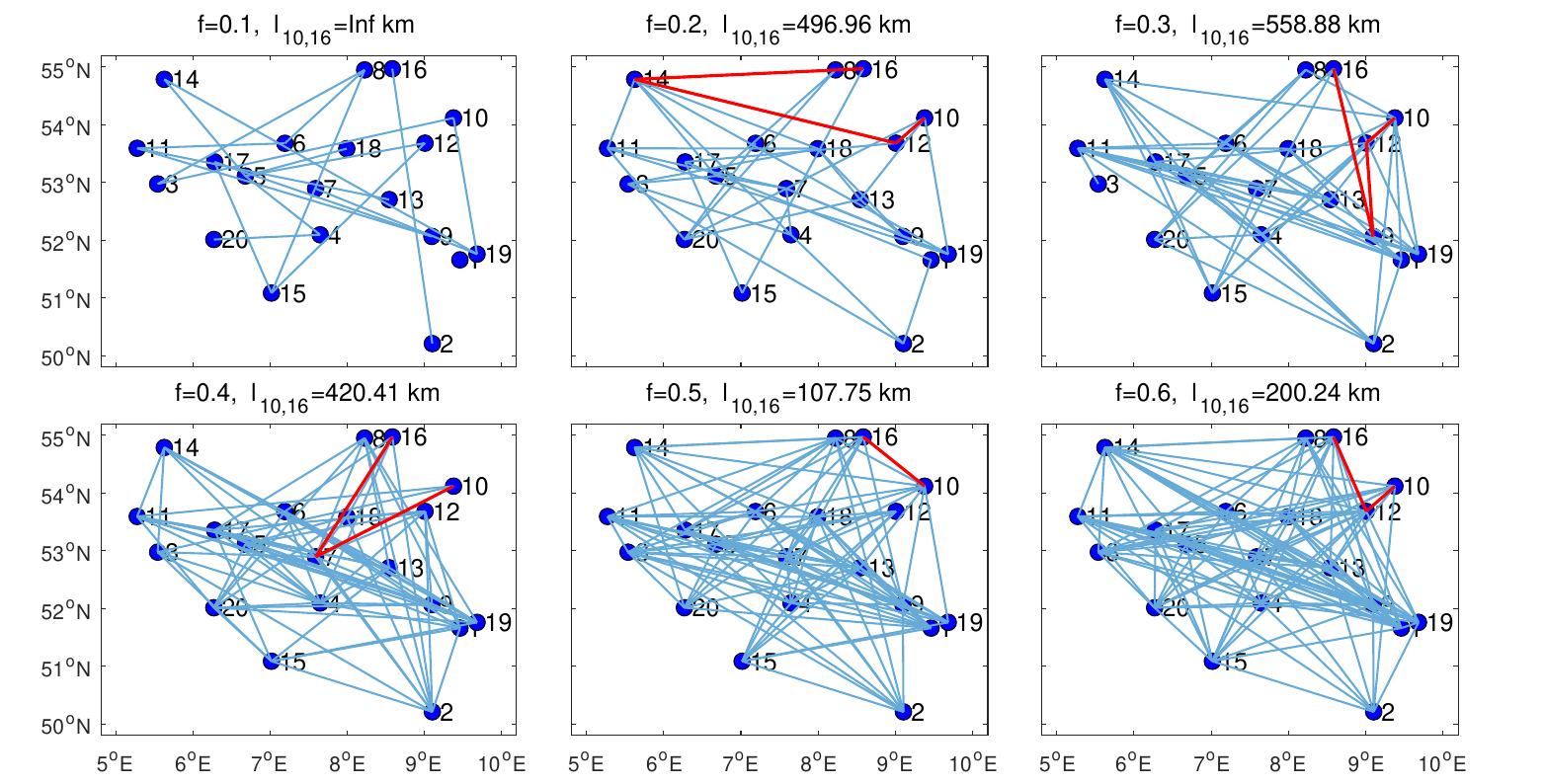}
\caption{Fully random networks with $n=20$ locations and connection fractions $f=0.1,0.2,\ldots,0.6$. The red lines highlight the shortest path between locations 10 and 16, where the network distance of this path is denoted by $l_{10,16}$.}
\label{figs4}
\end{center}
\end{figure*}

\begin{figure*}[tb!]
\begin{center}
\begin{overpic}[width=\textwidth]{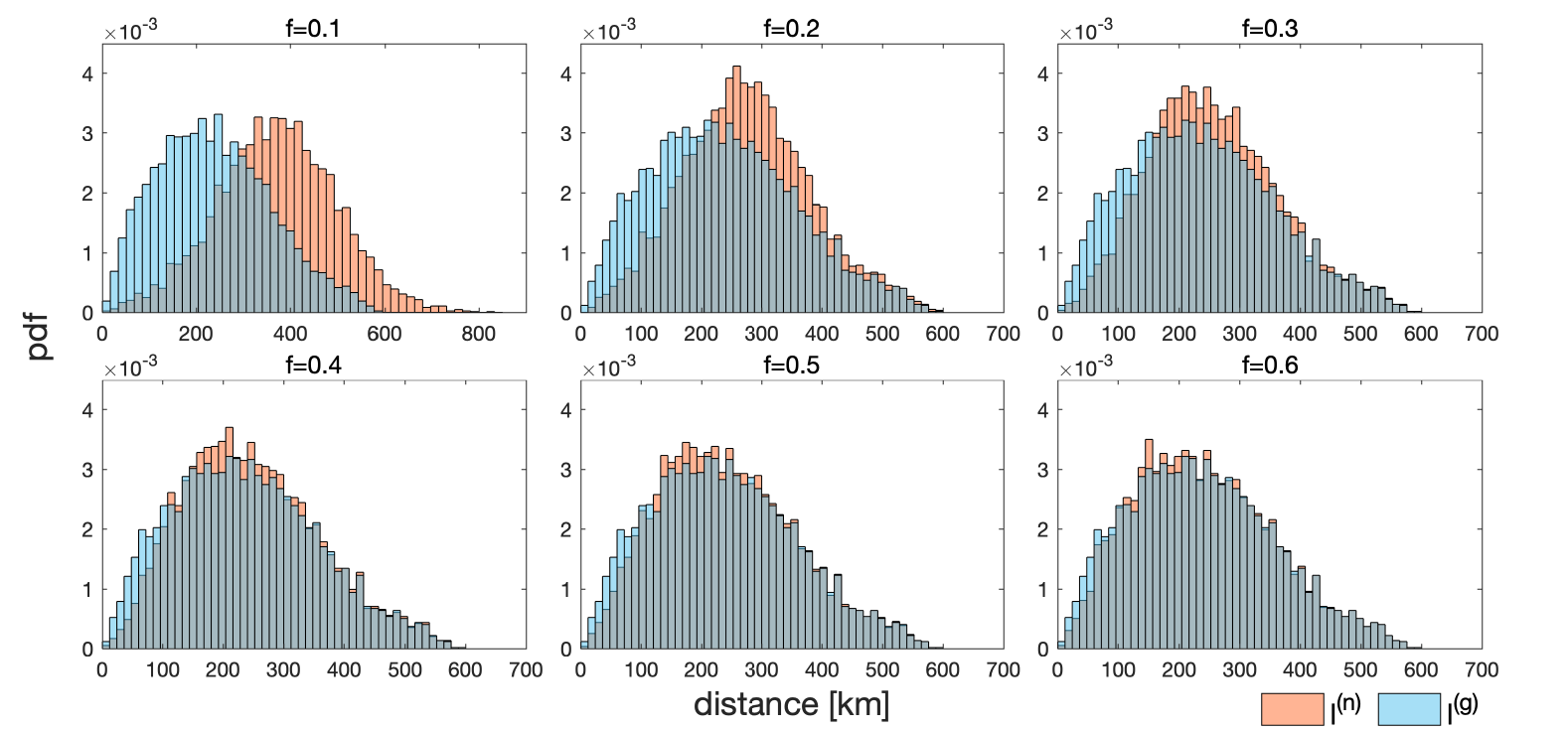}
\put(2,45){\bf \textsf a}
\end{overpic}
\\ \vspace{0.5cm}
\begin{overpic}[width=\textwidth]{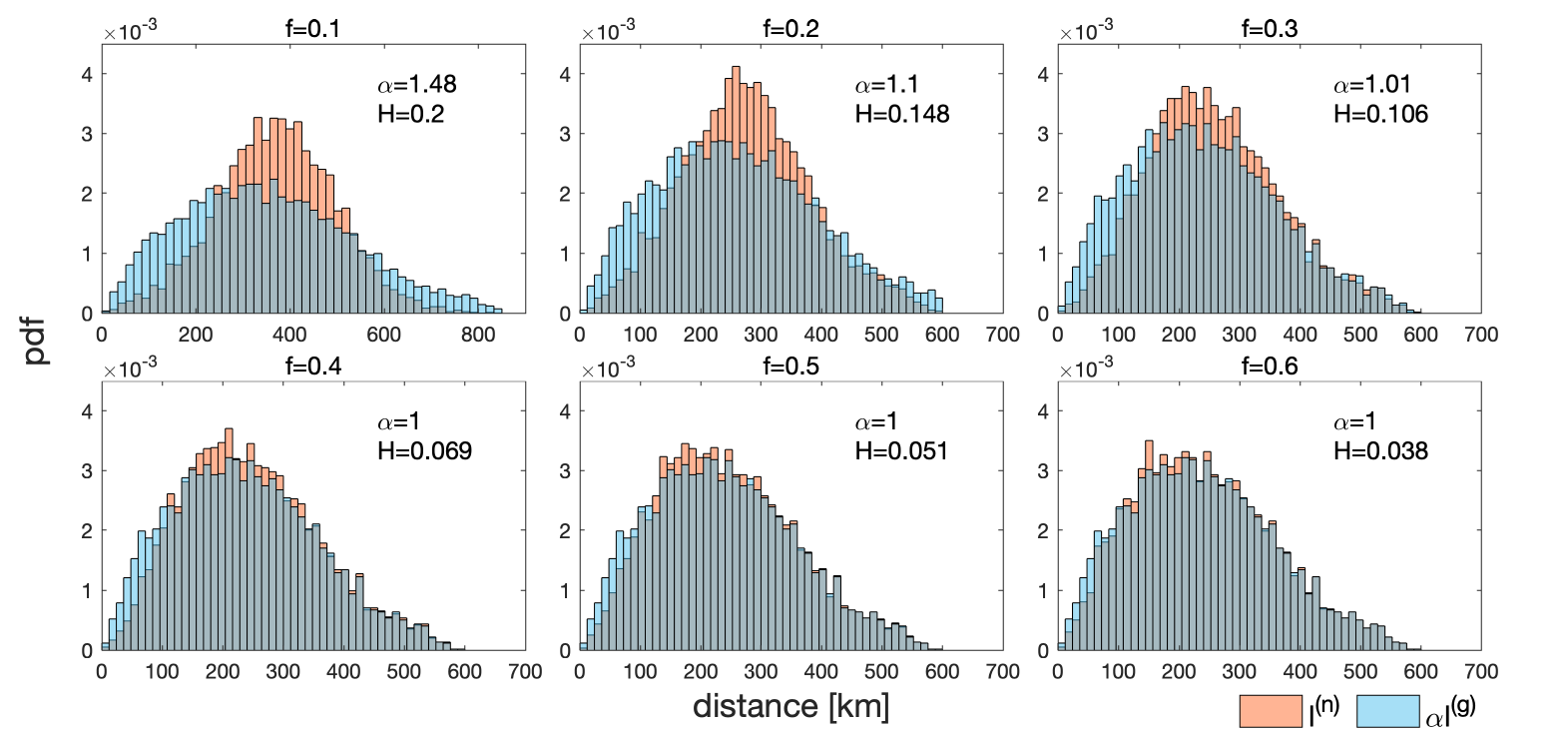}
\put(2,45){\bf \textsf b}
\end{overpic}
\caption{Distributions or pdfs $p^\text{(n)}(l)$ and $p^\text{(g)}(l)$ of network and geodetic distances for six fully random motorway networks with $n=100$ locations and connection fractions $f=0.1,0.2,\ldots,0.6$, before ({\textbf a}) and after ({\textbf b}) scaling. The values of the scaling factor $\alpha$ and the Hellinger distance $H$ are given in the subfigures of {\textbf b}.}
\label{figs5}
\end{center}
\end{figure*}

We randomly choose $n$ nodes as locations in the area between latitude 50$^\circ$ to 55$^\circ$ and longitude 5$^\circ$ to 10$^\circ$, i.e.~approximately $550~\text{km}\times 550~\text{km}$ in size, such that there are $N=n(n-1)/2$ possible motorways between them. For stipulated connection fractions $f=0.1,0.2,\ldots,0.6$, we randomly choose $fN$ motorways. In Fig.~\ref{figs4} we visualize this for $n=20$ locations. We notice that the network distance is the sum of the geodetic distances between the adjacent locations along the motorway path. If two locations are connected by a path in the motorway network, we work out the geodetic and the network distances. In the latter case, we pick the shortest one if there is more than one connecting path. For $f=0.01$ in the example of $n=20$ locations, see Fig.~\ref{figs4}, the path between locations 10 and 16 is absent and the network distance is infinite. For $n=100$ locations, as seen in Fig.~\ref{figs5}, the corresponding distributions strongly depend on the value of $f$. For smaller $f\leq 0.3$, the shapes are so different that scaling is absent, as revealed by the large Hellinger distances $H>0.1$. For $f\geq 0.4$ onwards, the distributions begin to coincide, and we trivially have $\alpha=1$. This behaviour does not match the empirical findings. In the real world, a motorway connecting two locations would not avoid a location in between and close by. Thus, the fully random motorway network contains a high number of unrealistic motorways, which alters the statistical features.

For empirical motorway networks, we find scaling factors $\alpha$ larger than one, which correspond to small $f$ for fully random networks with a large difference in distance distributions. This means empirical motorway networks cannot be described by a fully random network.

\vspace*{2cm}

\clearpage
\subsection{Random grid networks}

\begin{figure*}[htbp]
\begin{center}
\includegraphics[width=\textwidth]{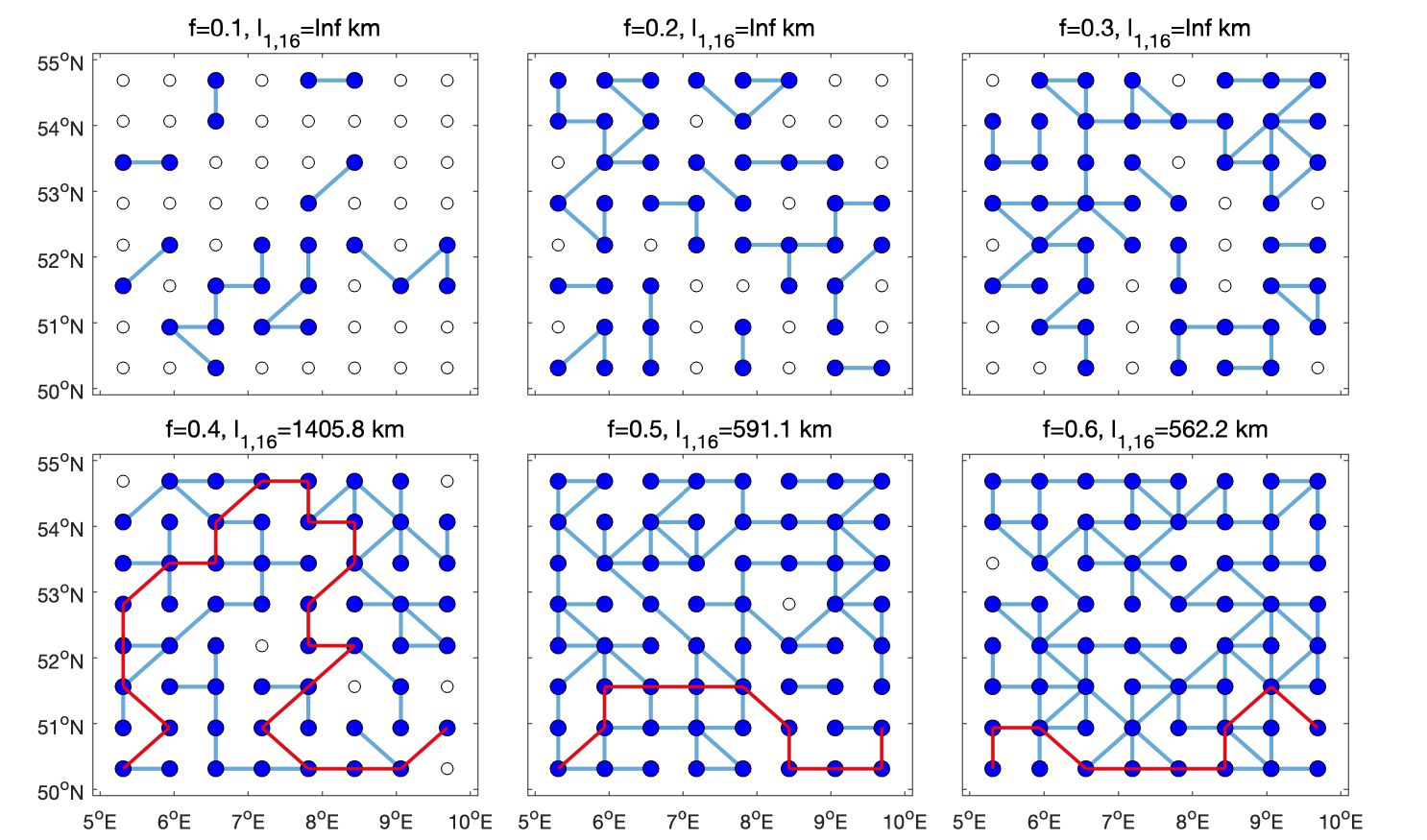}
\caption{Random grid networks with 64 locations on a $8\times 8$ grid and with connection fractions $f=0.1,0.2,\ldots,0.6$. The red lines highlight the shortest path between locations 1 and 16, where the network distance of this path is denoted by $l_{1,16}$.}
\label{figs6}
\end{center}
\end{figure*}

In the area between latitude 50$^\circ$ to 55$^\circ$ and longitude 5$^\circ$ to 10$^\circ$, we generate a fully connected graph with $xy$ locations in a $x\times y$ rectangle grid~\cite{Sohouenou2020}. All adjacent locations in one row/column are horizontally/vertically connected by motorways. The locations in odd rows and odd columns are diagonally connected with the adjacent ones in even rows and even columns. Therefore, there is no motorway crossing and the total number of connecting motorways is $(x-1)y+(y-1)x+(x-1)(y-1)$. If all possible motorways are realized, the network is fully connected. For connection fractions $f=0.1,0.2,\ldots,0.6$, we randomly select motorways. In Fig.~\ref{figs6}, we show examples for $x=y=8$, i.e.~for 64 locations. Network connectivity is usually not achieved for smaller $f$, since two locations may not be connected by a path in this random grid network. If a path exists, the network distance of the shortest path between two paired locations is likely to become shorter with increasing $f$, like the path between locations 1 and 16 in Fig.~\ref{figs6}. The network distance depends on the connection fraction and, certainly, on the number of motorways in the network.

\begin{figure*}[tb!]
\begin{center}
\begin{overpic}[width=\textwidth]{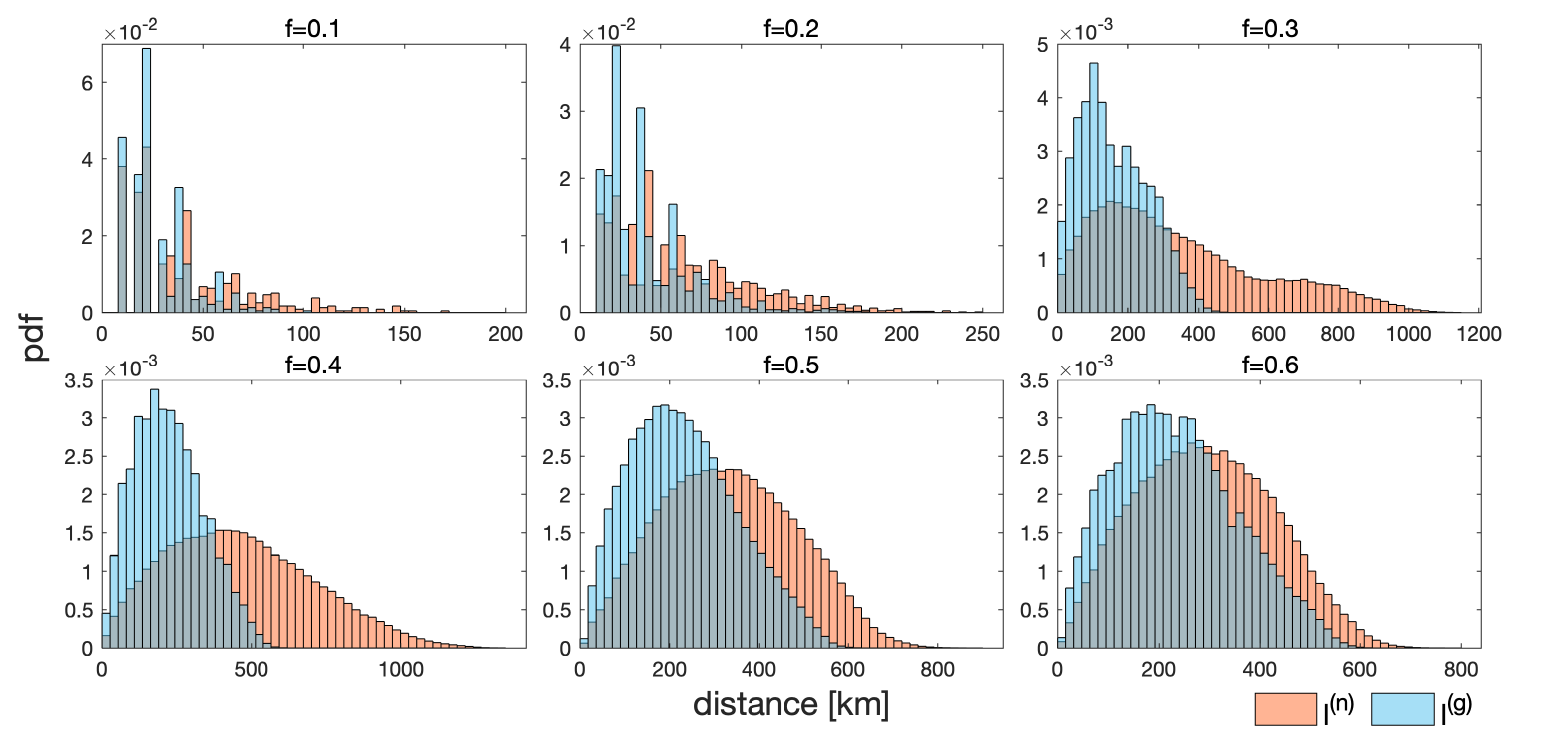}
\put(2,48){\bf \textsf a}
\end{overpic}
\\ \vspace{0.5cm}
\begin{overpic}[width=\textwidth]{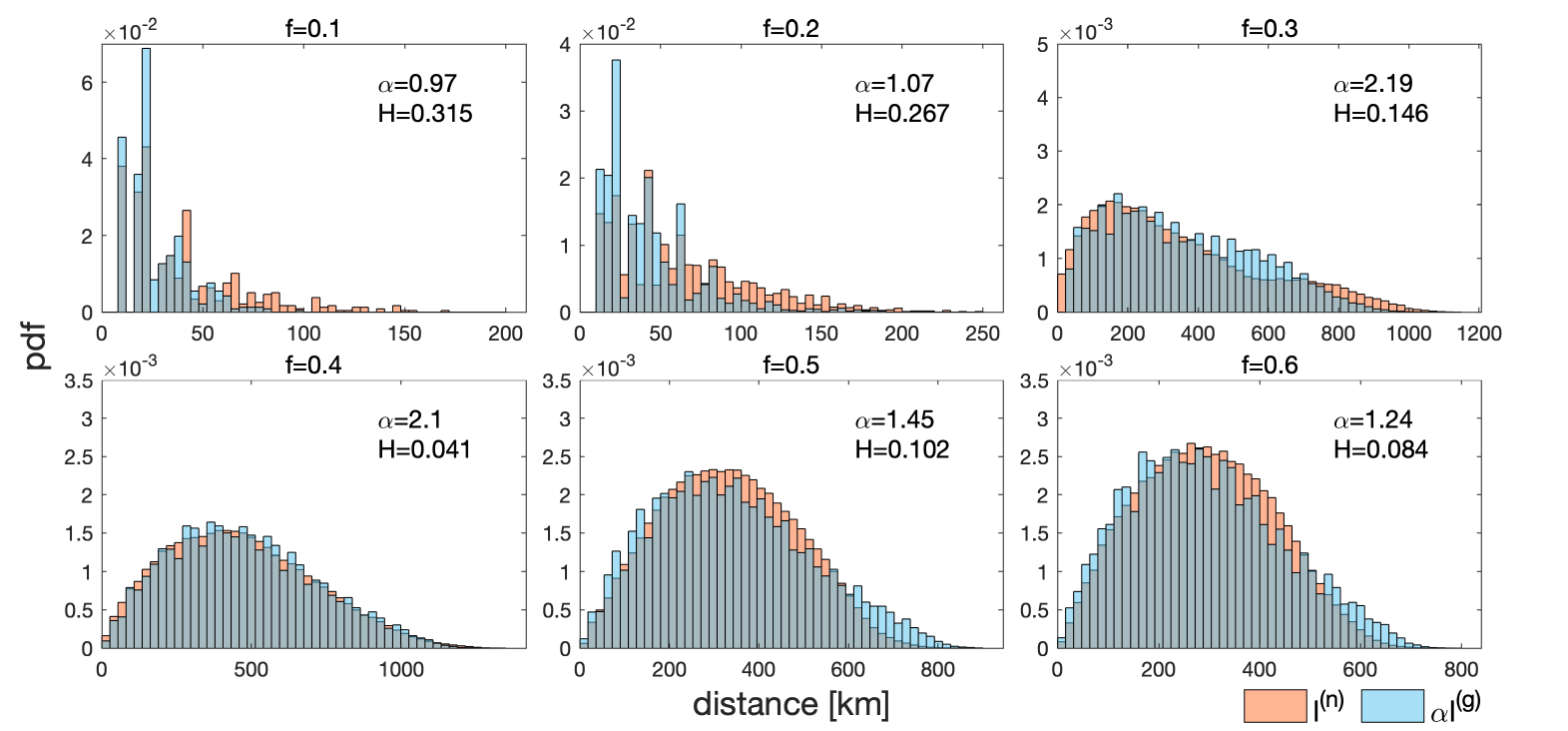}
\put(2,48){\bf \textsf b}
\end{overpic}
\caption{Distributions or pdfs $p^\text{(n)}(l)$ and $p^\text{(g)}(l)$ of network and geodetic distances for six random $30 \times 30$ grid motorway networks with $900$ locations and connection fractions $f=0.1,0.2,\ldots,0.6$, before ({\textbf a}) and after ({\textbf b}) scaling. The values of the scaling factor $\alpha$ and the Hellinger distance $H$ are given in the subfigures of {\textbf b}.}
\label{figs7}
\end{center}
\end{figure*}

We work out the corresponding distributions of network and geodetic distances for $x=y=30$, i.e.~for 900 locations. Again, the network distance is the sum of the geodetic distances along the path. The distributions in Fig.~\ref{figs7} are once more evidently different. For a small $f<0.4$, the distributions also deviate after scaling with large Hellinger distances $H>0.1$. For larger $f$, the distributions match after scaling. As $f$ increases, the number of motorways grows, and the network distances for a pair of locations tends to be smaller on average, lowering the scaling factor $\alpha$. Despite the emergence of the scaling behaviour in the random grid networks, the connection fraction of random grid networks is larger than $f=0.2214$ in the North Rhine-Westphalia region network, which is constructed to match the empirical situation. In the random grid networks for similar connection fractions $0.2\leq f\leq 0.3$, the scaling is poorly realized or absent.

\vspace*{2cm}

\clearpage
\subsection{Random geometric networks}

\begin{figure*}[htbp]
\begin{center}
\includegraphics[width=\textwidth]{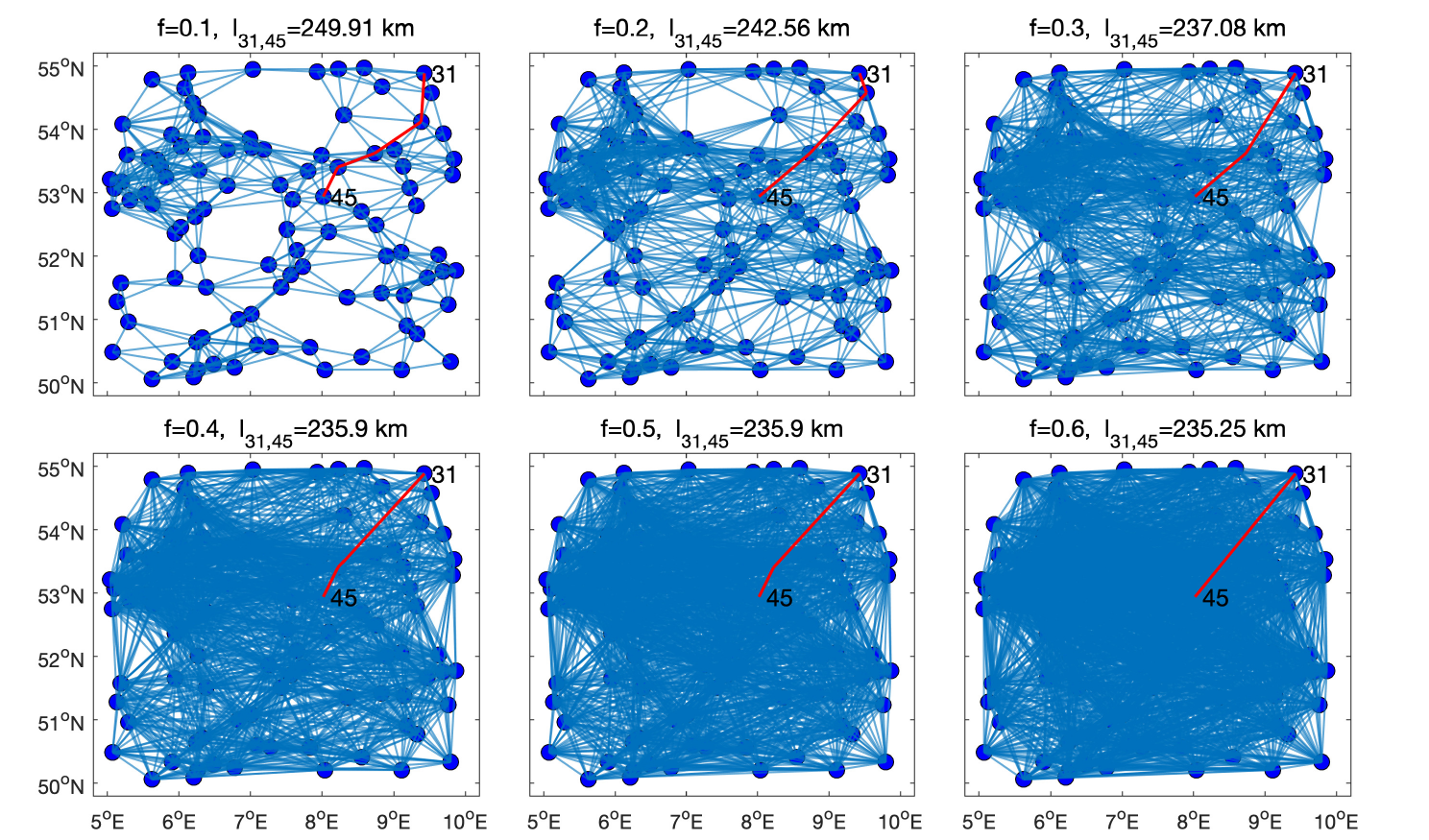}
\caption{Random geometric networks with $n=100$ locations and connection fractions $f=0.1,0.2,\ldots,0.6$. The red lines highlight the shortest path between locations 31 and 45, where the network distance of this path is denoted by $l_{31,45}$.}
\label{figs8}
\end{center}
\end{figure*}

\begin{figure*}[tb!]
\begin{center}
\begin{overpic}[width=\textwidth]{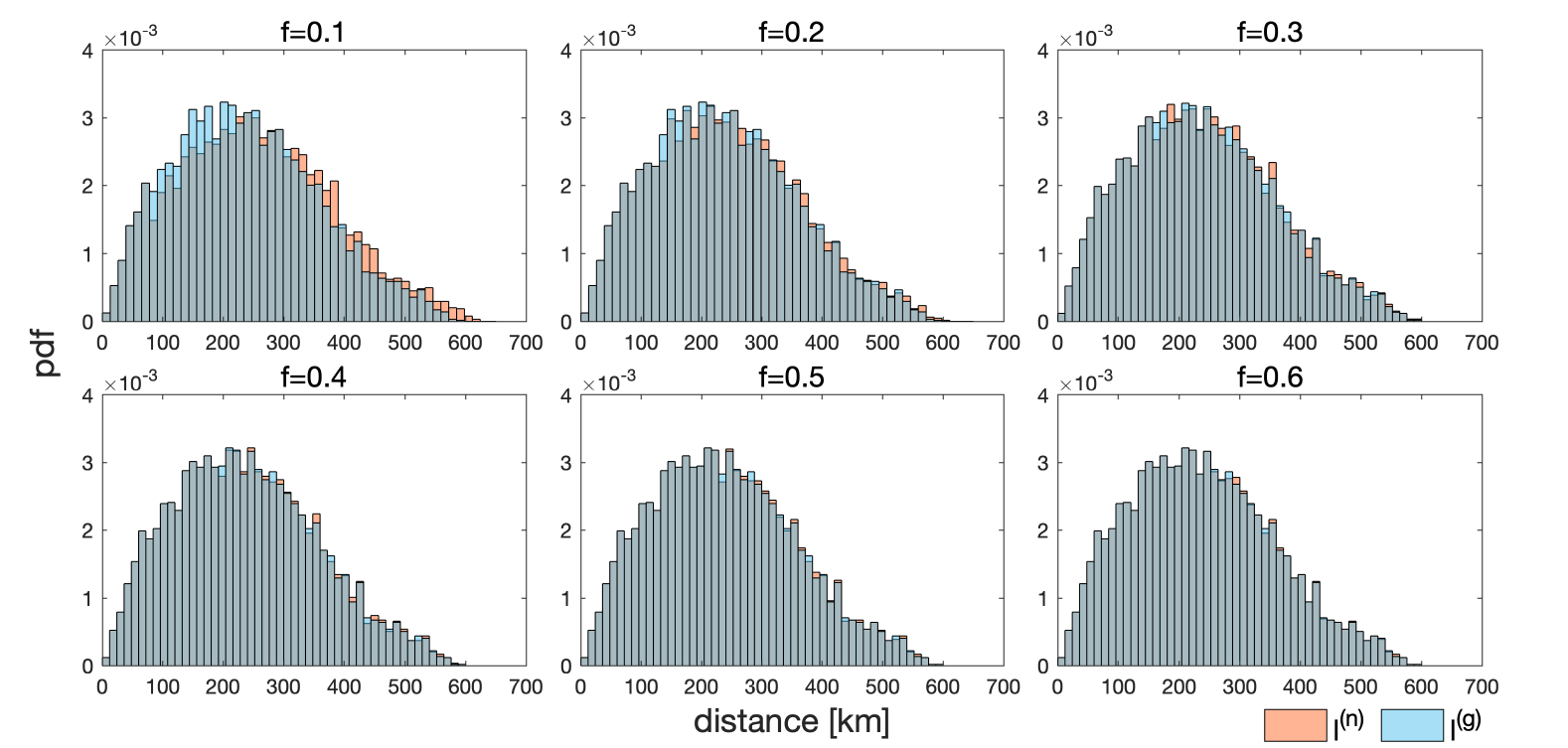}
\put(2,45){\bf \textsf a}
\end{overpic}
\\ \vspace{0.5cm}
\begin{overpic}[width=\textwidth]{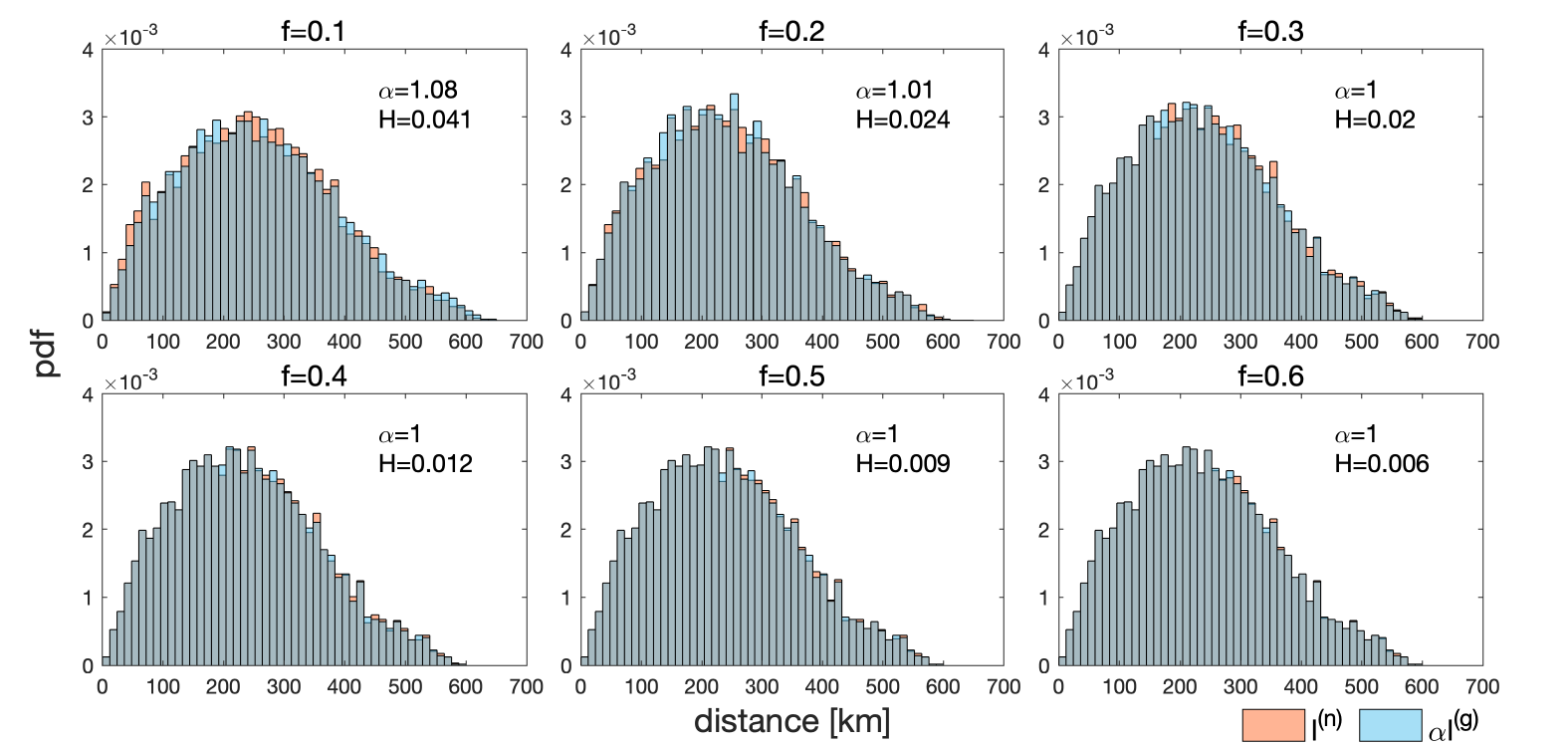}
\put(2,45){\bf \textsf b}
\end{overpic}
\caption{Distributions or pdfs $p^\text{(n)}(l)$ and $p^\text{(g)}(l)$ of network and geodetic distances for six random geometric motorway networks with $n=100$ locations and connection fractions $f=0.1,0.2,\ldots,0.6$, before ({\textbf a}) and after ({\textbf b}) scaling. The values of the scaling factor $\alpha$ and the Hellinger distance $H$ are given in the subfigures of {\textbf b}.}
\label{figs9}
\end{center}
\end{figure*}

In the area between latitude 50$^\circ$ to 55$^\circ$ and longitude 5$^\circ$ to 10$^\circ$, we randomly choose $n$ nodes as locations. If the geodetic distance between two locations $i$ and $j$ is less than or equal to a distance threshold $l_c$ ($0<l_c<\infty$)~\cite{Aldous2010}, i.e., $l_{ij}^{(g)}\leq l_c$ with $i\neq j$, we connect the two locations by a motorway. In this manner, we construct a random geometric motorway network. Here the distance threshold contributes to the locally finite configurations. The number of connections divided by $N=n(n-1)/2$ possible motorways yields the connection fraction $f$ of the network. Taking 100 random locations into account, we vary the distance threshold, which results in different random geometric networks as well as different connection fractions. Figure~\ref{figs8} displays six generated networks with $f=0.1, 0.2,\cdots,0.6$. The red line in each subfigure highlights the path between locations 31 and 45, which becomes shorter and more straight with $f$ increasing. This implies that the increasing availability of paths make the network distance close to a straight-line distance or geodetic distance. The distributions of network and geodetic distances for the six random geometric motorway networks in Fig.~\ref{figs9} reveal little difference before the scaling. Therefore, after the scaling, the factor $\alpha$ is either close to or equal to 1 with a rather small Hellinger distance. These cases, in particular the case of $f=0.2$, are rather different from the above analysed real motorway networks, suggesting the locality in network connections is not the main feature of a realistic motorway network.

\vspace*{2cm}

\clearpage
\subsection{$K$-neighbour networks}

\begin{figure*}[htbp]
\begin{center}
\includegraphics[width=\textwidth]{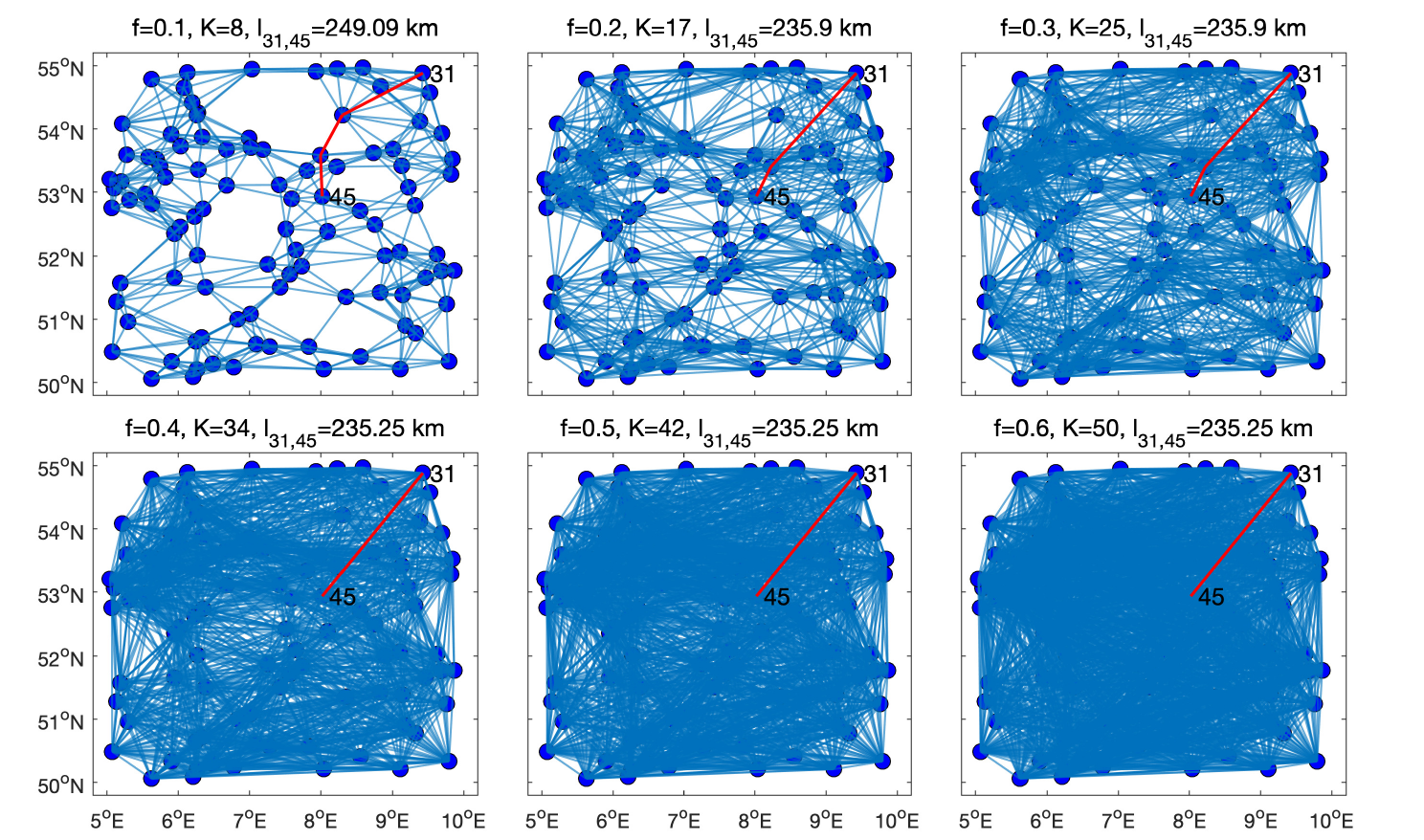}
\caption{$K$-neighbour networks with $n=100$ locations and connection fractions $f=0.1,0.2,\ldots,0.6$. The red lines highlight the shortest path between locations 31 and 45, where the network distance of this path is denoted by $l_{31,45}$.}
\label{figs10}
\end{center}
\end{figure*}

A $K$-neighbour motorway network features the locality of network connections as well but in a distinct configuration of the random geometric motorway networks. We first randomly choose $n$ locations in the area between latitude 50$^\circ$ to 55$^\circ$ and longitude 5$^\circ$ to 10$^\circ$. For a pair of locations $i$ and $j$, if location $i$ is one of the $K$ closest neighbours of location $j$ or location $j$ is one of the $K$ closest neighbours of location $i$~\cite{Aldous2010}, we connect the two locations. Increasing $K$ from 1 to $n-1$ yields different $K$-neighbour motorway networks and the corresponding connection fractions $f$. The networks for $f=0.1, 0.2, \cdots, 0.6$ shown in Fig.~\ref{figs10} are similar to but not exactly the same as the corresponding ones of the random geometric motorway networks in Fig.~\ref{figs8}. The distributions of network and geodetic distances are indistinguishable for $f>0.1$ before scaling and the factor $\alpha$ approaches to 1 after scaling. Analogously to the random geometric motorway networks, the scaling behavior in distance distributions obviously deviates from the case of real motorway networks. This once more demonstrates the locality of network connections cannot fully describe a realistic motorway network, which uses a very different configuration.

\vspace*{2cm}

\begin{figure*}[tb!]
\begin{center}
\begin{overpic}[width=\textwidth]{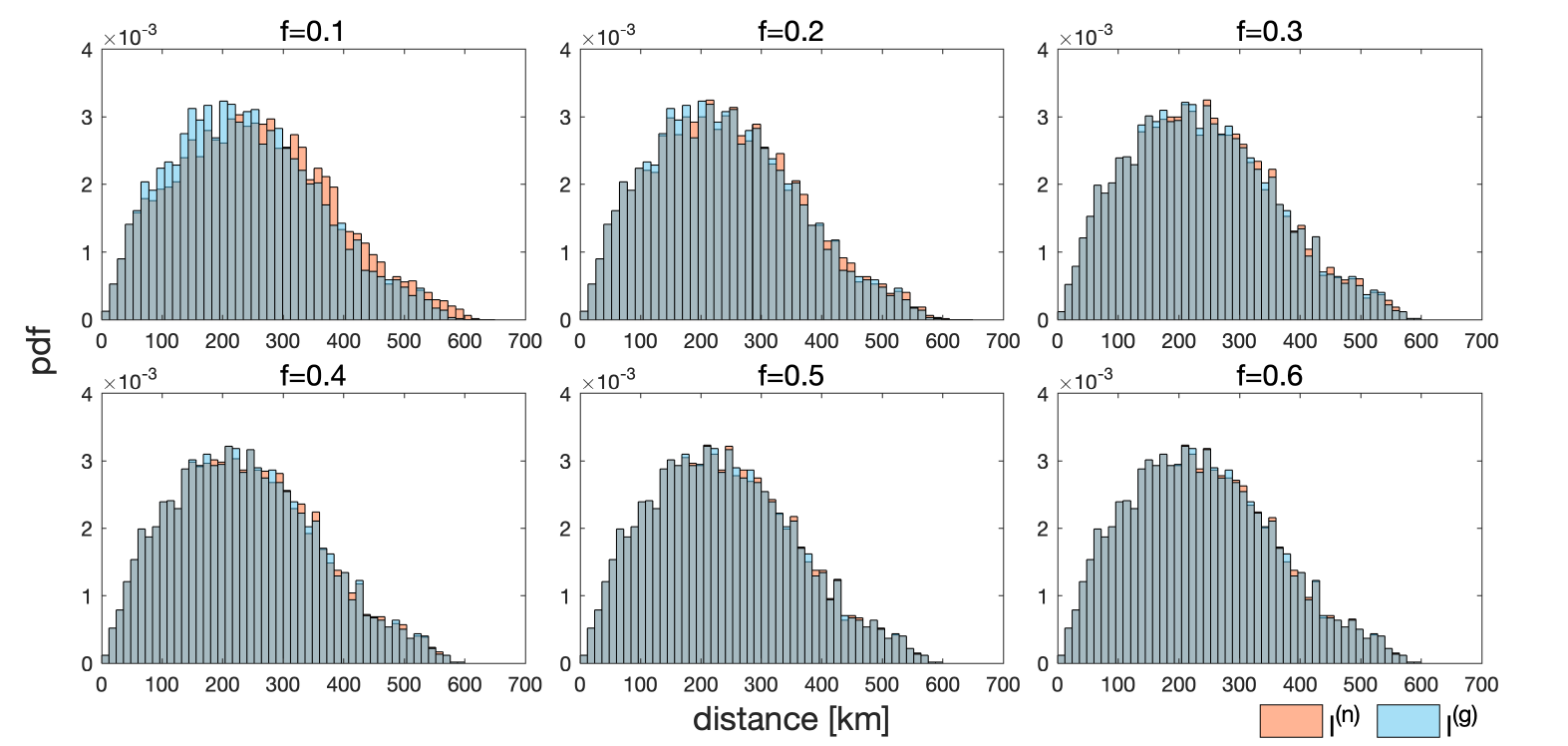}
\put(2,45){\bf \textsf a}
\end{overpic}
\\ \vspace{0.5cm}
\begin{overpic}[width=\textwidth]{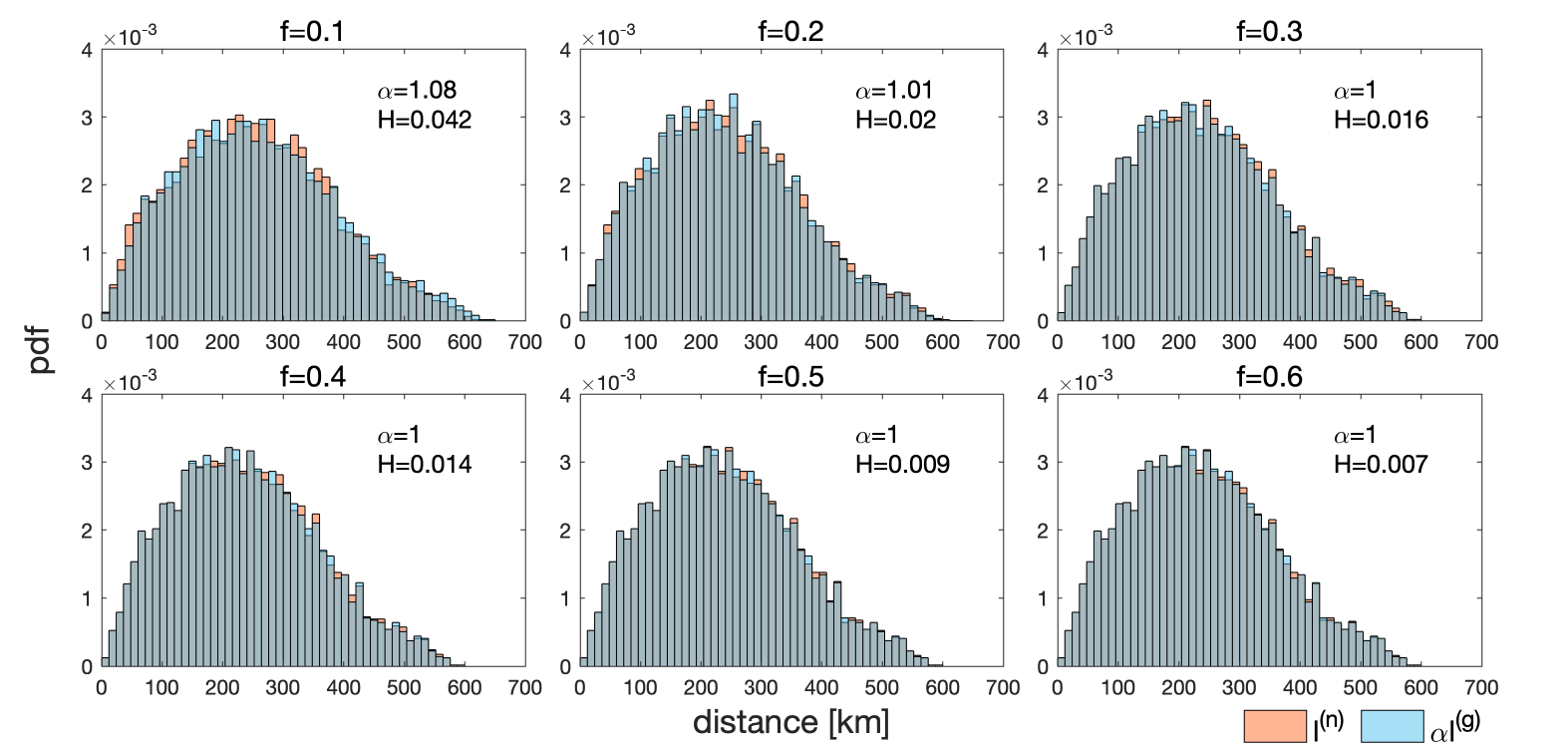}
\put(2,45){\bf \textsf b}
\end{overpic}
\caption{Distributions or pdfs $p^\text{(n)}(l)$ and $p^\text{(g)}(l)$ of network and geodetic distances for six $K$-neighbour motorway networks with $n=100$ locations and connection fractions $f=0.1,0.2,\ldots,0.6$, before ({\textbf a}) and after ({\textbf b}) scaling. The values of the scaling factor $\alpha$ and the Hellinger distance $H$ are given in the subfigures of {\textbf b}.}
\label{figs11}
\end{center}
\end{figure*}

\clearpage
\section{Modelled motorway networks}

\begin{figure*}[htbp]
\begin{center}
\includegraphics[width=\textwidth]{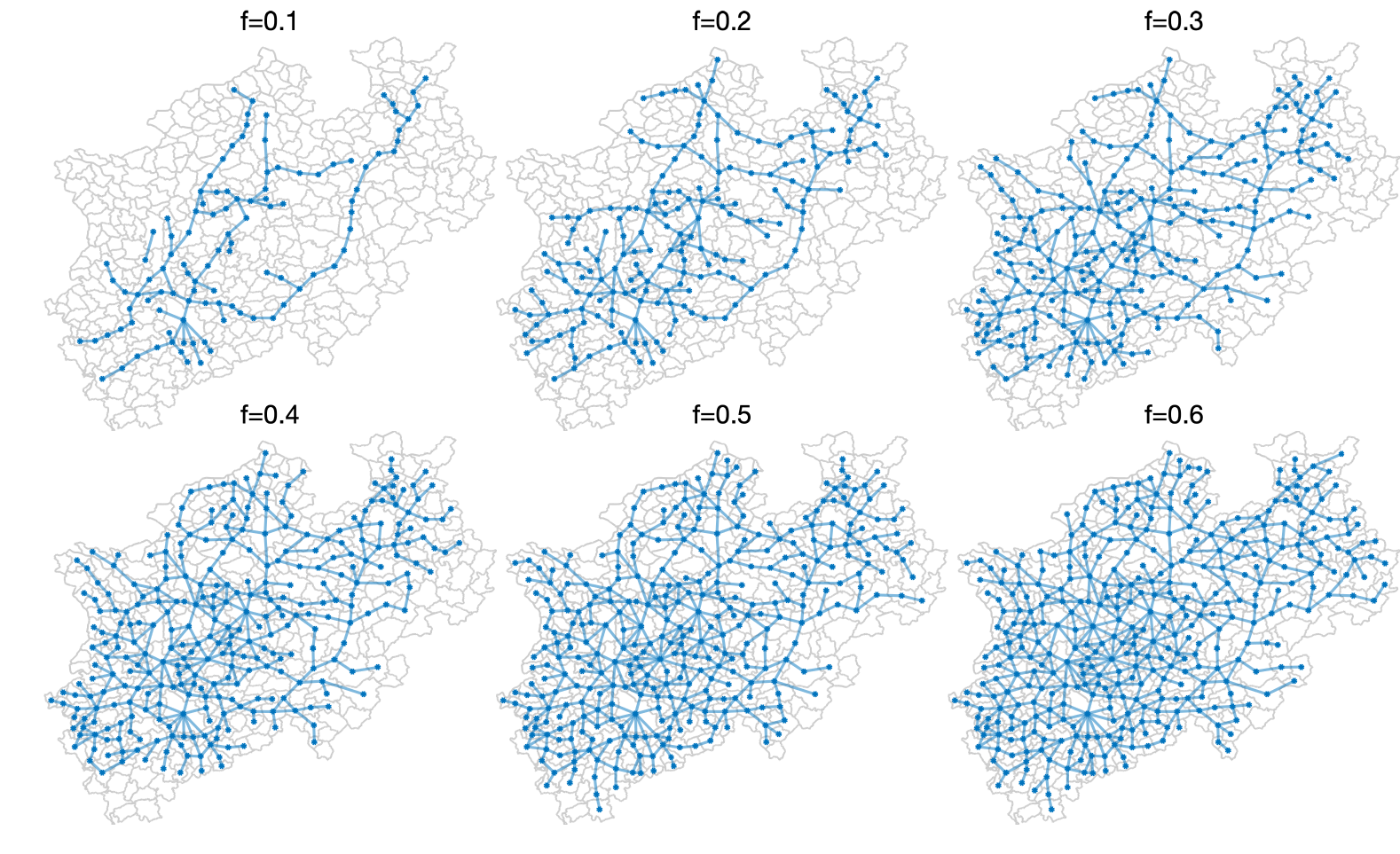}
\caption{Partially random motorway networks with connection probabilities $f=0.1$, 0.2, 0.3, 0.4, 0.5, 0.6 for the North Rhine-Westphalia region network. Maps developed by Matlab~\cite{matlab}.}
\label{figs12}
\end{center}
\end{figure*}

\begin{figure*}[htbp]
\begin{center}
\begin{overpic}[width=\textwidth]{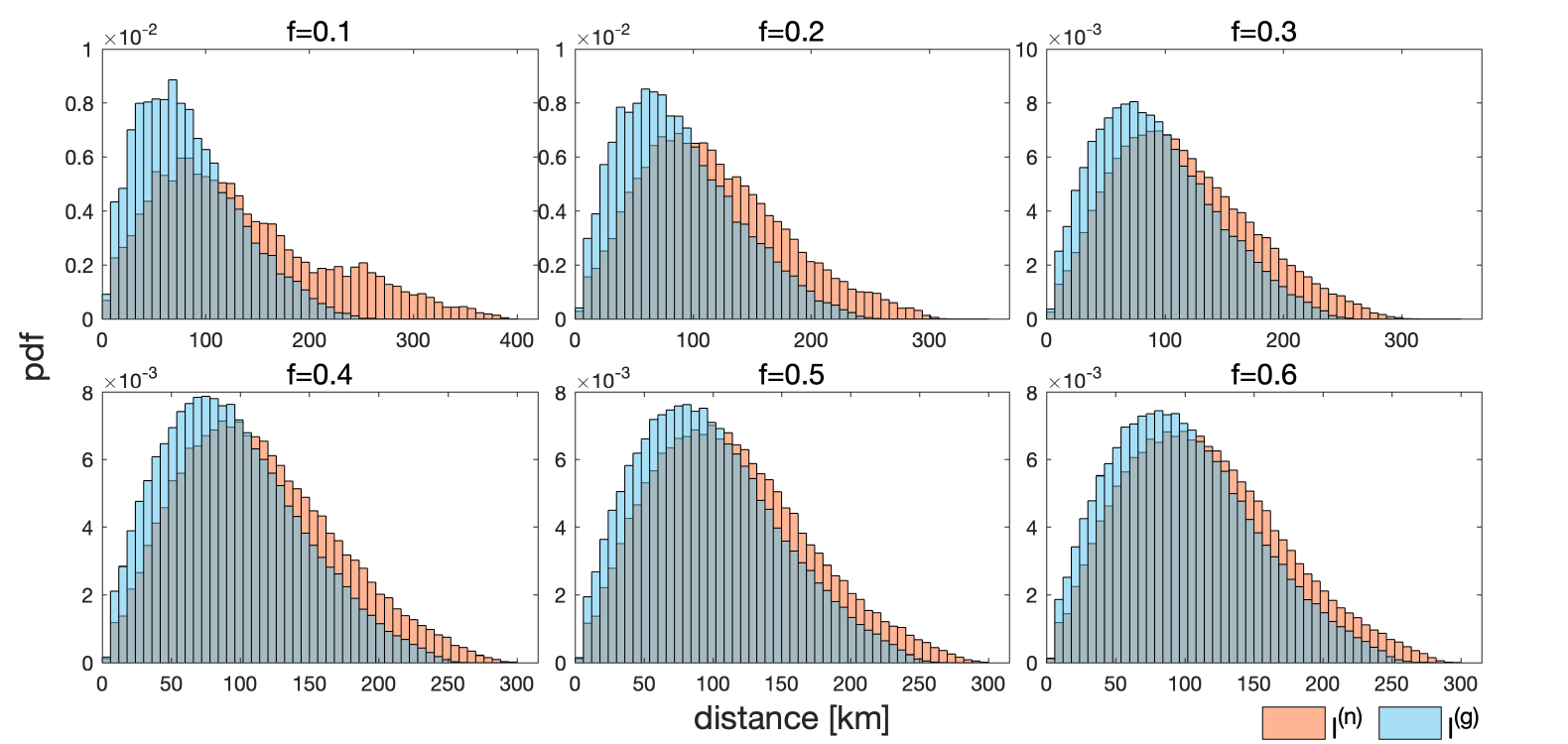}
\put(2,45){\bf \textsf a}
\end{overpic}
\\ \vspace{0.5cm}
\begin{overpic}[width=\textwidth]{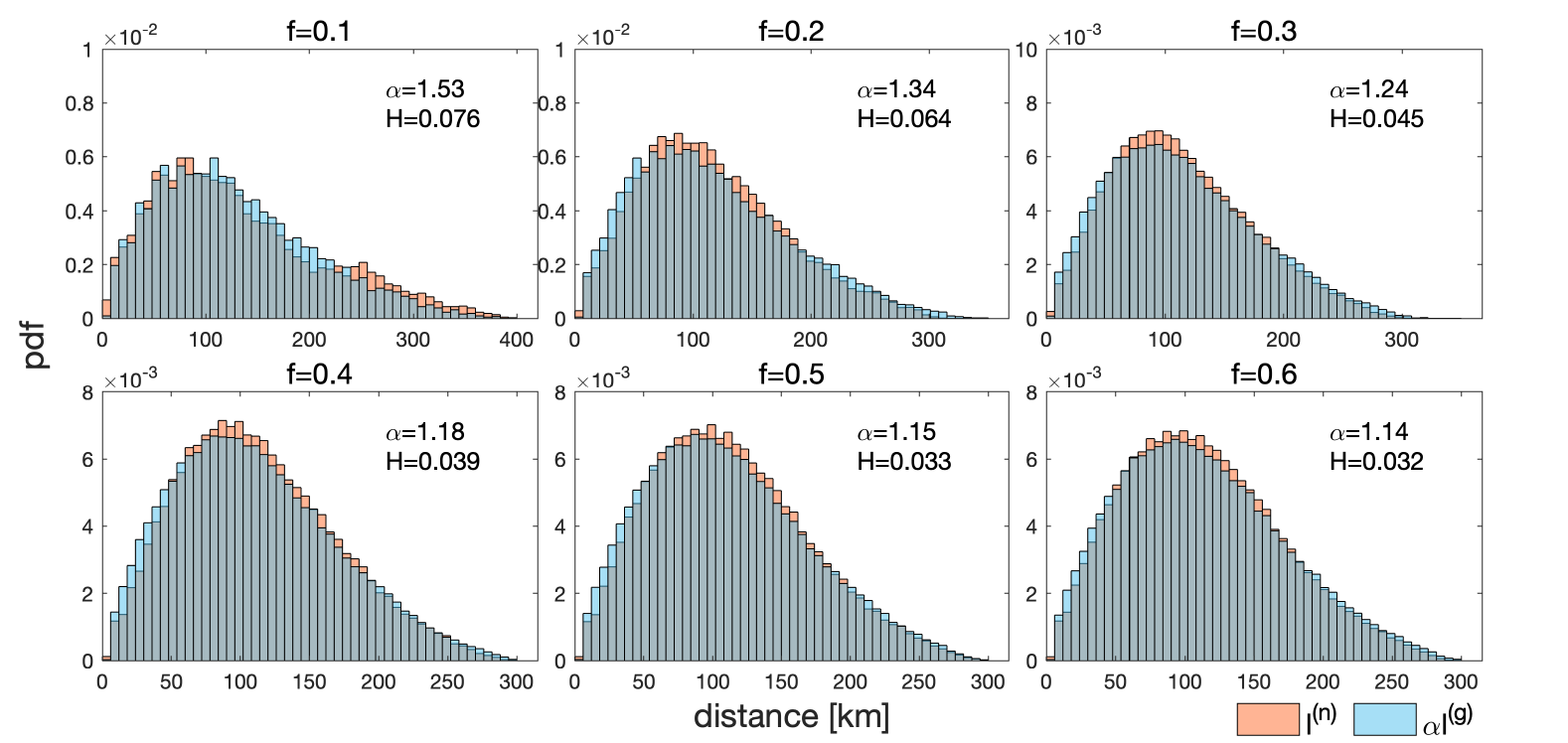}
\put(2,45){\bf \textsf b}
\end{overpic}
\caption{Distributions or pdfs $p^\text{(n)}(l)$ and $p^\text{(g)}(l)$ of network and geodetic distances for the partially random motorway networks shown in Fig.~\ref{figs12} before ({\textbf a}) and after ({\textbf b}) scaling. Scaling factors $\alpha$ and Hellinger distances $H$ are given in the subfigures of {\textbf b}.}
\label{figs13}
\end{center}
\end{figure*}

\begin{figure*}[htbp]
\begin{center}
\includegraphics[width=\textwidth]{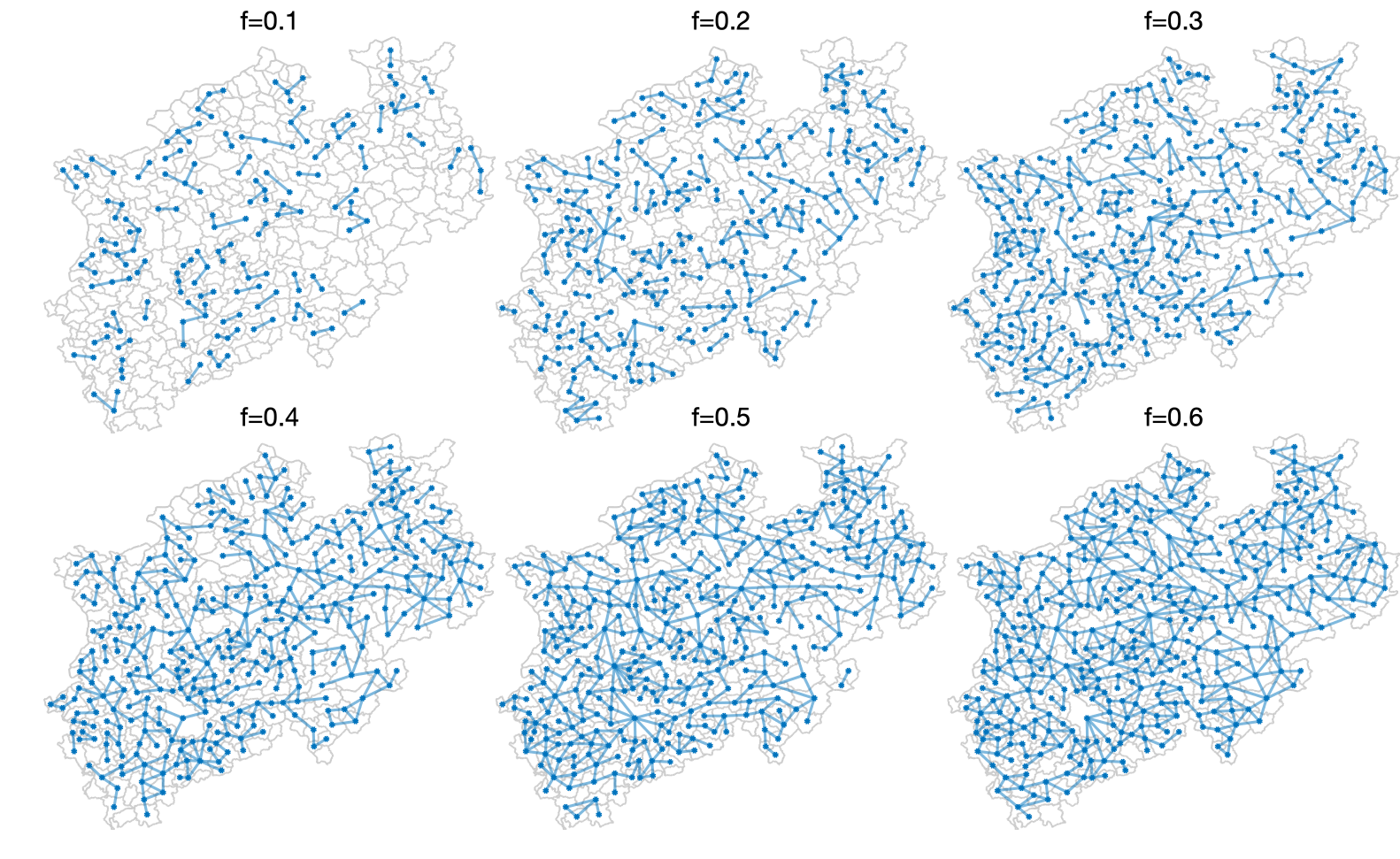}
\caption{Fully random motorway networks with connection probabilities $f=0.1$, 0.2, 0.3, 0.4, 0.5, 0.6 for the North Rhine-Westphalia region network. Maps developed by Matlab~\cite{matlab}.}
\label{figs14}
\end{center}
\end{figure*}

\begin{figure*}[htbp]
\begin{center}
\begin{overpic}[width=\textwidth]{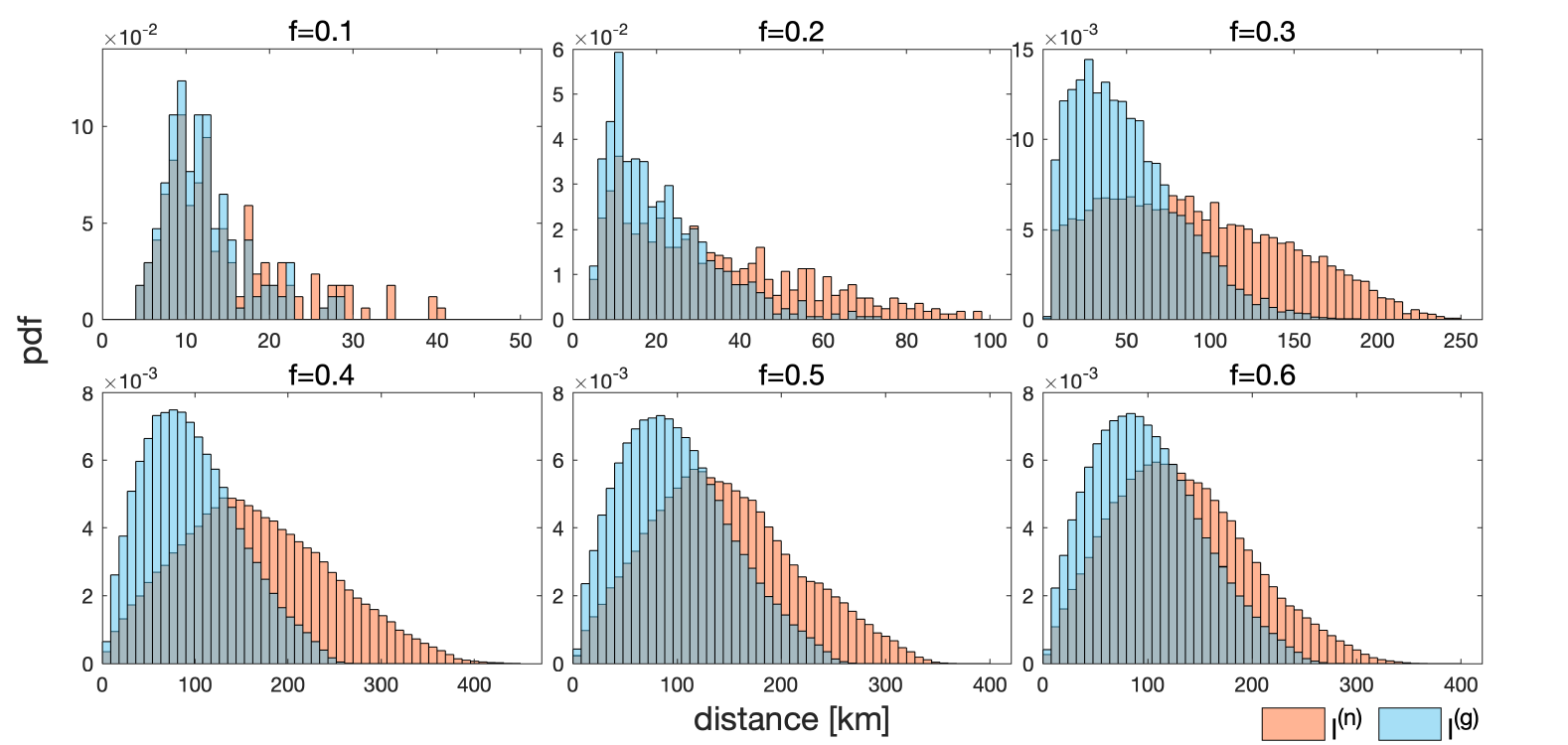}
\put(2,48){\bf \textsf a}
\end{overpic}
\\ \vspace{0.5cm}
\begin{overpic}[width=\textwidth]{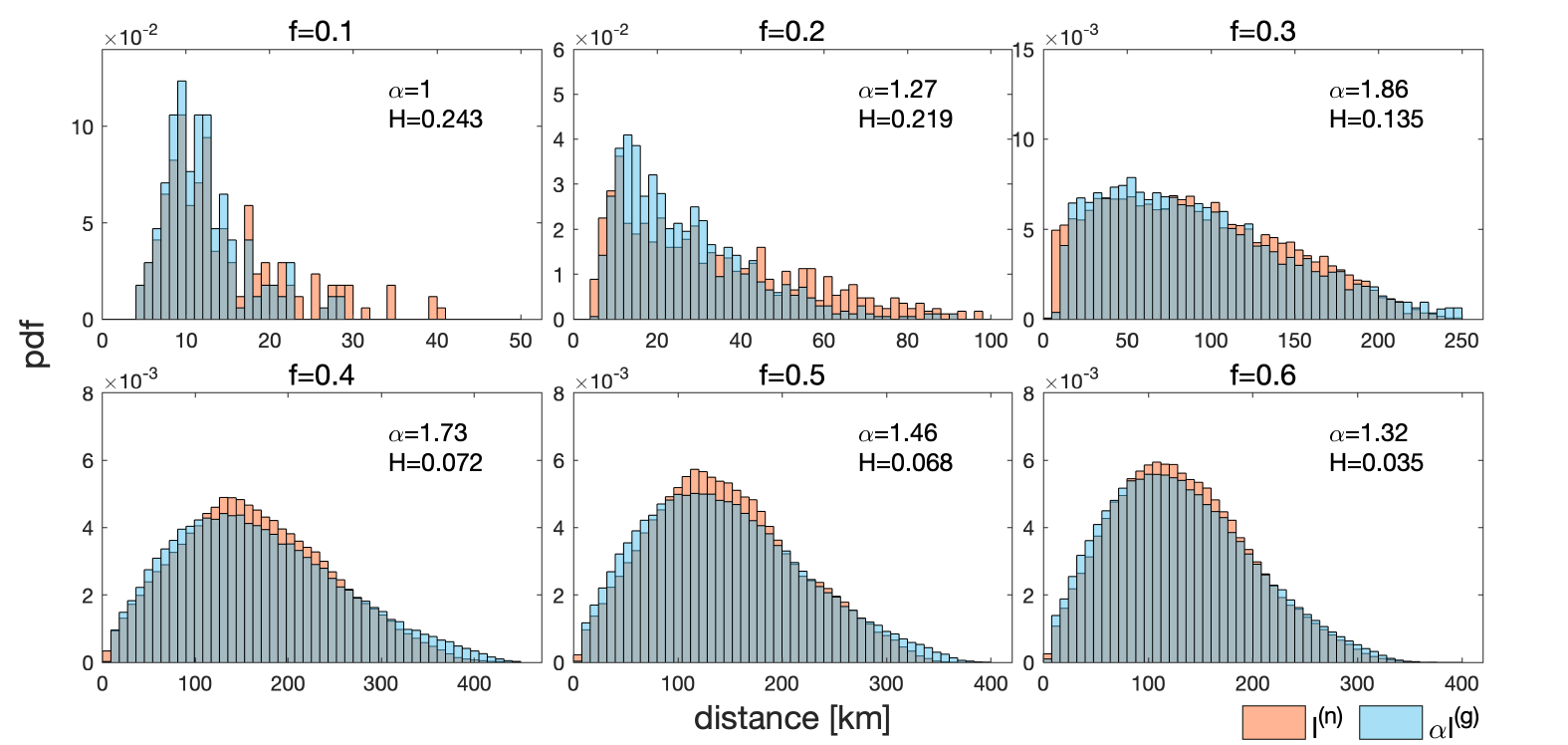}
\put(2,48){\bf \textsf b}
\end{overpic}
\caption{Distributions or pdfs $p^\text{(n)}(l)$ and $p^\text{(g)}(l)$ of network and geodetic distances for the fully random motorway networks shown in Fig.~\ref{figs14} before ({\textbf a}) and after ({\textbf b}) scaling. Scaling factors $\alpha$ and Hellinger distances $H$ are given in the subfigures of {\textbf b}.}
\label{figs15}
\end{center}
\end{figure*}

In the paper, we describe the modelling of partially random motorway networks and the fully random motorway networks based on North Rhine-Westphalia. Here we present the modelled partially random motorway networks for different connection fractions $f$, $f=0.1$, 0.2, 0.3, 0.4, 0.5, 0.6, in Fig.~\ref{figs12}. For comparison, we show fully random motorway networks for the same connection fractions $f$ in Fig.~\ref{figs14}. We display the distributions $p^\text{(n)}(l)$ and $p^\text{(g)}(l)$ of network and geodetic distances in Figs.~\ref{figs13} and~\ref{figs15} for the partially and fully random motorway networks, respectively.

\vspace{3cm}

\clearpage
\section{A simple, qualitative model}

A simple, qualitative model helps to clarify the shape of the distance distributions. It does not explain the scaling. Here, we consider the unimodal case and make the important assumption that the network covers the area $A$ densely such that any distance $l$ can be found on the network. It thus suffices to work with a smooth distribution $f(\vec{r})$ of locations $\vec{r}\in A$. Furthermore, we neglect curvature effects, i.e. we view $A$ as planar. The approximation for the distribution $p^{(\text{g})}(l)$ then reads
\begin{equation}
p^{(\text{approx})}(l)=\int\limits_A d^2 r_1\int\limits_A d^2 r_2 f(\vec{r}_1) f(\vec{r}_2) \delta (l-|\vec{r}_1-\vec{r}_2|)\ .
\label{eqs51}
\end{equation} 
As an example, shown in Fig.~\ref{figs12}a, we look at a disk-shaped area with radius $R$ centred at the origin, $A=C_R(0)$, and a homogeneous distribution
\begin{equation}
f(\vec{r})=\frac{1}{\pi R^2} \Theta (|\vec{r}-R|)
\label{eqs52}
\end{equation}
of locations. Here, $\Theta$ is the Heaviside function. The resulting distribution $p^{(\text{approx})}(l)$ is shown in Fig.~\ref{figs12}b. Its shape does not resemble our empirical findings. We now argue that this is due to the assumption of a homogeneous distribution~\eqref{eqs52}, which is often unrealistic. As seen in Fig.~1 in the paper, the motorway networks in countries such as China, Germany and the contiguous US show higher densities inside the area than close to the borders. We thus assume a Gaussian distribution
\begin{equation}
f(\vec{r}|\sigma)=\frac{1}{2\pi \sigma^2} \exp \left(-\frac{\vec{r}^2}{2\sigma^2}\right)
\label{eqs53}
\end{equation}
of locations where the standard deviation $\sigma$ should be estimated by the sample standard deviation. According to the fall-off of the distribution, the boundaries of the area $A$ considered become less important, and we extend the integration over the entire plane. A simple calculation yields
\begin{equation}
p^\text{approx}(l)=\frac{1}{2\sigma^2}l\exp\left(-\frac{l^2}{4\sigma^2}\right)
\label{eqs54}
\end{equation}
We fit this distribution to the empirical results of the geodetic distributions for China, France, Germany and North Rhine-Westphalia. The fits in Fig.~\ref{figs9} agree surprisingly well, given the simplicity of the model~\eqref{eqs51} and the Gaussian assumption~\eqref{eqs53}. This is corroborated by the numerical values for the standard deviations $\sigma$, the only fit parameter, which are rather close to the ones obtained by sampling for the geodetic distances. This simple model captures the almost linear increase near the origin, the asymmetry and the fall-off of the empirical distributions for the geodetic distances. The corresponding distribution $p^\text{(n)}(l)$ for the network distances in this simple model follows from Eq.~(1) in the paper.

\begin{figure*}[htb!]
\begin{center}
\hspace*{-1cm}
\begin{overpic}[width=\textwidth]{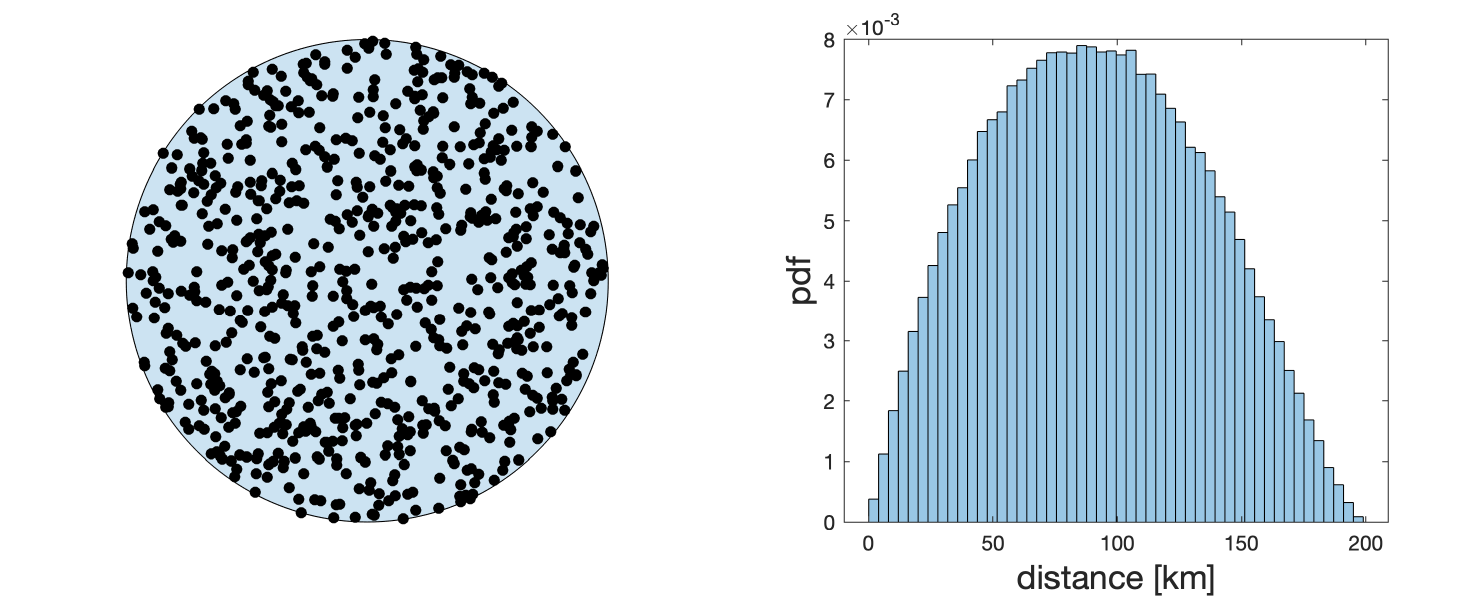}
\put(5,38){{\bf \textsf a}}
\put(53,38){{\bf \textsf b}}
\end{overpic}
\caption{{\textbf a}, 774 locations uniformly distributed in a disk with a radius of 100 km. {\textbf b}, distributions of geodetic distances, where geodetic distances are equal to Euclidean distances, as the disk is planar without curvature effects.}
\label{figs16}
\end{center}
\end{figure*}

\begin{figure*}[ht!]
\begin{center}
\includegraphics[width=\textwidth]{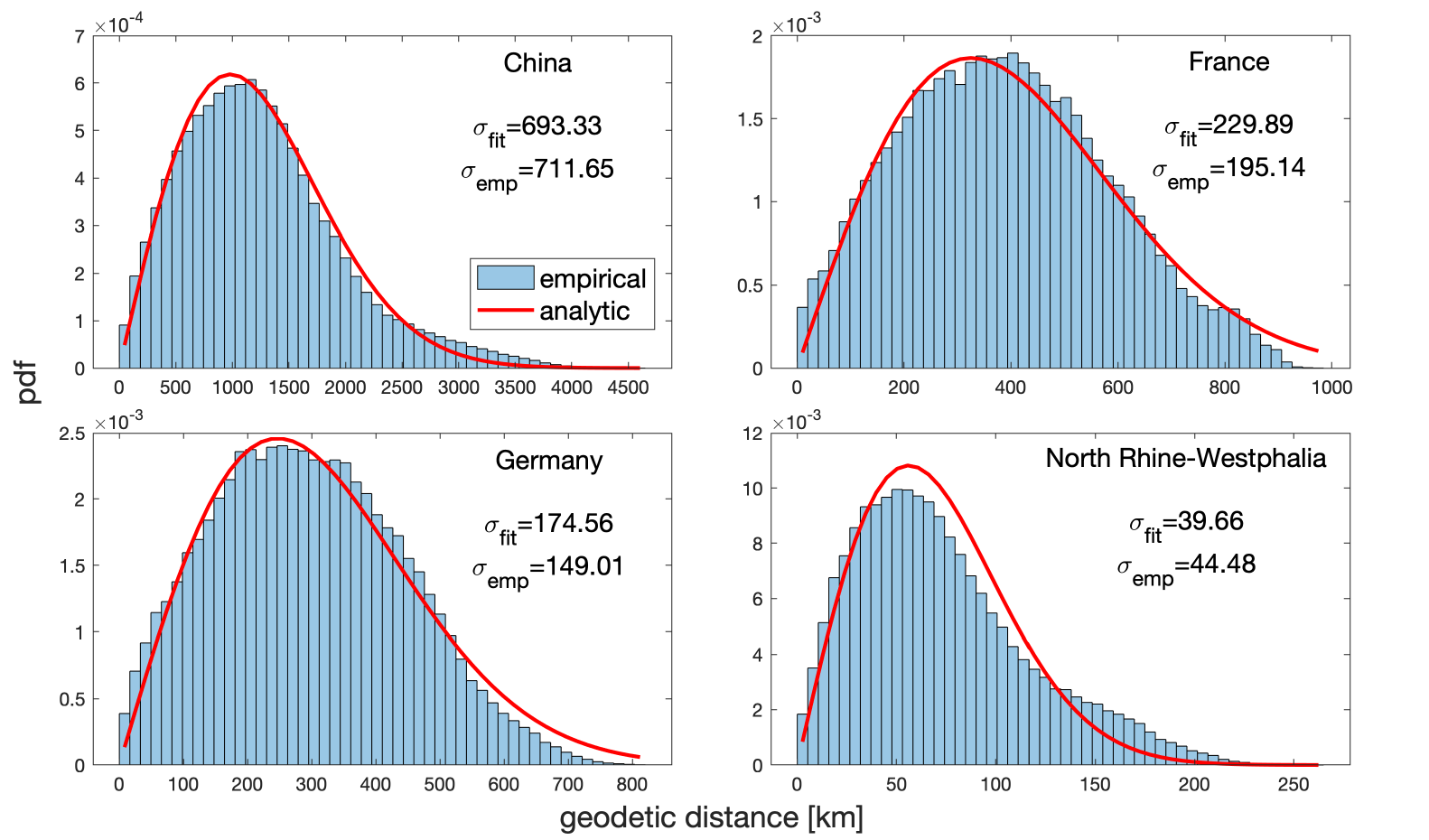}
\caption{Distributions of geodetic distances for China, France, Germany and North Rhine-Westphalia, fitted with analytic equation~\eqref{eqs54}. Empirical standard deviations $\sigma_\text{emp}$ and fitted standard deviations $\sigma_\text{fit}$ of geodetic distances are given in each subfigure.}
\label{figs17}
\end{center}
\end{figure*}